\documentclass[11pt]{article}
\usepackage[margin=1.1in]{geometry}
\usepackage[numbers,sort&compress]{natbib}

\usepackage{graphicx}
\usepackage{hyperref}
\usepackage{float}
\usepackage{epstopdf}
\usepackage{amsmath}
\usepackage{verbatim}
\usepackage[english]{babel}
\usepackage{multirow}
\renewcommand{\arraystretch}{1.3} % to make tables more readable
\usepackage{array}
\usepackage{tabularx}
\usepackage{longtable}

\newcolumntype{P}[1]{>{\raggedright\arraybackslash}p{#1}}
\newcolumntype{Q}{>{\raggedright\arraybackslash}X}

\newcommand{\MateraTableSetup}{\small\setlength{\tabcolsep}{6pt}\renewcommand{\arraystretch}{1.2}}
\newcommand{\MateraTH}[1]{\textbf{#1}}

\usepackage[utf8]{inputenc}
 
\usepackage{listings}
\usepackage{xcolor}
 
\definecolor{codegreen}{rgb}{0,0.45,0.15}
\definecolor{codegray}{rgb}{0.75,0.75,0.75}
\definecolor{codepurple}{rgb}{0.2235,0.1608,0.4941}
\definecolor{codeblue}{rgb}{0.0,0.2,0.6}
\definecolor{codebrown}{rgb}{0.6,0.2,0.0}
\definecolor{codeblack}{rgb}{0.15,0.15,0.15}
\definecolor{backcolour}{rgb}{0.98,0.98,0.98}

\lstdefinestyle{mystyle}{
    language=Python,
    backgroundcolor=\color{backcolour},
    basicstyle=\scriptsize\ttfamily\color{codeblack},
    commentstyle=\color{codegreen}\itshape,
    keywordstyle=\color{magenta}\bfseries,
    numberstyle=\tiny\color{codegray},
    stringstyle=\color{codebrown},
    identifierstyle=\color{codeblack},
    emph={Material,Materials,ZSLInterfaceAnalyzer,create_interface,create_slab},
    emphstyle=\color{codepurple}\bfseries,
    breaklines=true,
    captionpos=b,
    keepspaces=true,
    numbers=left,
    numbersep=4pt,
    showstringspaces=false,
    xleftmargin=0.1\linewidth,
    xrightmargin=0.1\linewidth,
    frame=single,
    framerule=0pt,
    rulecolor=\color{backcolour},
    framesep=2pt,
    aboveskip=10pt,
    belowskip=10pt,
    columns=fullflexible,
    tabsize=2
}
\title{AI-ready design of realistic 2D materials and \\ interfaces with Mat3ra-2D}

\author{
    Vsevolod Biryukov\thanks{Exabyte Inc. (Mat3ra.com), Walnut Creek, CA 94596, USA}
    \and
    Kamal Choudhary
        \thanks{Department of Materials Science and Engineering, Johns Hopkins University, Baltimore, MD 21218, USA} 
        \thanks{Department of Electrical and Computer Engineering, Johns Hopkins University, Baltimore, MD 21218, USA}
    \and
    Timur Bazhirov\footnotemark[1] \footnotemark[4]
}
\date{}

\begin{document}
    \maketitle
    \begingroup
    \renewcommand{\thefootnote}{\fnsymbol{footnote}}
    \footnotetext[4]{Corresponding author: timur@mat3ra.com}
    \endgroup
    \begin{abstract}
    Artificial intelligence (AI) and machine learning (ML) models in materials science are predominantly trained on ideal bulk crystals, limiting their transferability to real-world applications where surfaces, interfaces, and defects dominate. We present Mat3ra-2D, an open-source framework for rapid design of realistic two-dimensional materials and related structures, including slabs and heterogeneous interfaces, with support for disorder and defect-driven complexity. The approach combines: (1) well-defined standards for storing and exchanging materials data with a modular implementation of core concepts and (2) transformation workflows expressed as configuration-builder pipelines that preserve provenance and metadata. We implement typical structure generation tasks, such as constructing orientation-specific slabs or strain-matching interfaces, in reusable Jupyter notebooks that serve as both interactive documentation and templates for reproducible runs. To lower the barrier to adoption, we design the examples to run in any web browser and demonstrate how to incorporate these developments into a web application. Mat3ra-2D enables systematic creation and organization of realistic 2D- and interface-aware datasets for AI/ML-ready applications.
    
\end{abstract}

    \section{Introduction}
\label{sec:introduction}

    % Importance of data in materials science
    Modern materials research increasingly relies on digital data and automation to shorten the time from hypothesis to validated insight. Large-scale open databases and workflow management systems have enabled high-throughput computational screening and accelerated discovery across broad application domains \cite{jain2013materialsproject, curtarolo2012aflowlib, saal2013openQMD, pizzi2016aiida, nomad,choudhary2020joint,lee2025agapi}. At the same time, the processes that govern materials properties in practice combined with the rapid growth of atomically thin materials has created a need for datasets and tooling that go beyond ideal bulk crystals and capture the structures most relevant to experiments and devices, including monolayers, slabs, interfaces and heterostructures with defects and disorder\cite{haastrup2018computational, rasmussen2015computational,mathew2016mpinterfaces,choudhary2017high,choudhary2023efficient,choudhary2024intermat}.
    
    A persistent limitation is that many machine-learning models and benchmark datasets are dominated by idealized three-dimensional crystals, while real-world performance often depends on surfaces, reconstructions, disorder, and defects, as well as heterogeneous interfaces that determine transport, catalysis, and stability \cite{ward2016ml-framework, isayev2017ml-descriptors}. Bridging this gap requires (i) systematic generation of realistic structures and (ii) a consistent representation of structure, transformations, and provenance suitable for databases and AI/ML pipelines.
    
    This challenge is particularly acute for 2D and surface-dominated systems. In such cases, the relevant scientific degrees of freedom often arise from how a structure is created rather than from stoichiometry alone: orientation, termination, layer count, vacuum spacing, strain state, stacking registry, defect type, and post-processing steps can all alter the resulting material model. As a result, generating realistic structures for computation is not simply a matter of selecting a material from a database, but of constructing and documenting a sequence of transformations that leads to the target geometry. As demonstrated in the recent study\cite{choudhary2025chipsff}, comparing 16 graph-based machine-learned force fields, the lack of data on structural features relevant to real-world semiconductor applications can hinder the model's applicability to interfaces.
    
    Mat3ra-2D addresses these needs with an open-source framework for practical design of realistic 2D materials and related structures, including slabs, interfaces, defects, and other low-dimensional motifs relevant to device modeling. The framework combines shared data standards, reusable transformation logic, provenance-aware workflows, and example-driven access paths so that realistic structures can be generated, inspected, reused, and adapted across different computational settings. In this sense, it builds on our prior work on accessible computational materials design \cite{2018-exabyte-accessible-CMD}, data-centric ecosystem design \cite{EB-ARX-02-2019}, practical categorization of computational models \cite{EB-ARX-2021}, interpretable materials ML \cite{EB-ARX-2023}, and, most directly, M-CODE as a categorization and ontology layer for realistic structures \cite{biryukov2026mcode}.

    Mat3ra-2D is part of Mat3ra.com, a collaborative platform for computational materials R\&D \cite{EB-ARX-02-2019}, and is focused on organizing realistic structure generation as a reusable software stack spanning standards, common data, implementation packages, and interactive notebooks. The approach is designed to interoperate with established Python tooling for materials analysis and atomistic modeling, including pymatgen and ASE \cite{ong2013python, larsen2017atomic}, while also supporting browser-based execution for lower-friction access to representative workflows. The resulting focus on explicit structure-generation choices is aligned with our broader effort to make materials datasets and models more traceable and reusable across the digital materials workflow \cite{EB-ARX-2021, EB-ARX-2023, biryukov2026mcode}.

    The emphasis of the present work is therefore practical as well as conceptual. We focus not only on what kinds of realistic structures can be represented, but also on how they can be created through reusable workflows, documented with explicit provenance, and shared through openly accessible notebooks. This combination is intended to support both systematic dataset generation and more transparent communication of how realistic computational structures are built.

    The remainder of this manuscript is organized as follows. Section~\ref{sec:methodology} outlines the overall approach and the main components of the ecosystem. Section~\ref{sec:results} summarizes representative outcomes and example workflows enabled by Mat3ra-2D. Section~\ref{sec:discussion} discusses implications, practical considerations, and future directions. Section~\ref{sec:conclusion} concludes with a brief summary.

    \section{Methodology}
\label{sec:methodology}
    \subsection{Overview}
    \label{subsec:methodology-overview}

        Mat3ra-2D is organized as an implementation stack for practical generation of realistic structures, such as slabs, surfaces, interfaces, and defective configurations, with explicit provenance. Rather than presenting a single monolithic codebase, the methodology is distributed across standards, common data, implementation layers, and usage examples, each with a distinct role in making realistic structure generation reusable and reproducible.

    \subsection{Components}
    \label{subsec:methodology-components}

        \subsubsection{Standards}
        \label{subsubsec:methodology-components-standards}

            The lowest layer of the stack is the standards package \texttt{mat3ra-esse}~\cite{mat3ra-esse-pypi-package, mat3ra-esse-npm-package}, which provides the schemas, ontology, categorization, and exchange format used throughout the ecosystem. Its role is to provide a shared machine-readable representation for structures, inputs, and provenance so that higher-level tools operate on consistent concepts.

            These standards, together with representative schema and JSON examples, are based on the M-CODE categorization and ontology described in the corresponding manuscript \cite{biryukov2026mcode}. In the present work, we use that foundation as the substrate for realistic 2D structure generation rather than reintroducing the full categorization itself.

        \subsubsection{Commonly-used data}
        \label{subsubsec:methodology-components-standata}

            The next layer is \texttt{mat3ra-standata}~\cite{mat3ra-standata-pypi-package}, which distributes commonly used example materials and related records that adhere to the shared standards. In practice, this package serves as a reusable source of input structures for workflows and notebooks, allowing examples to begin from consistent reference materials rather than redefining them locally.

        \subsubsection{OOD abstraction layer}
        \label{subsubsec:methodology-components-code}

            Above the standards layer, \texttt{mat3ra-code} provides implementations of the abstract level of the standardized entities as Python and JavaScript/TypeScript classes. This approach is versed in the object-oriented design (OOD). This layer is not the main user-facing entry point for the present work; instead, it acts as an abstraction layer used by higher-level packages. Its role is to keep the common "abstract" portions of the specific concepts (such as materials, configurations, and transformations, described in the subsequent section) consistent across the codebase and to reduce duplication when the same concepts are used in different tools.

        \subsubsection{Materials design implementation}
        \label{subsubsec:methodology-components-made}

            The main implementation layer is \texttt{mat3ra-made}~\cite{mat3ra-made-pypi-package}. This package contains the concrete materials-design functionality used to construct realistic structures: material objects, builders, analyzers, and modifiers for operations such as slab generation, interface construction, strain handling, and defect creation. In practical terms, this is the layer where standardized inputs become executable workflows.

\begin{figure}
                \centering
                \includegraphics[width=1.0\textwidth]{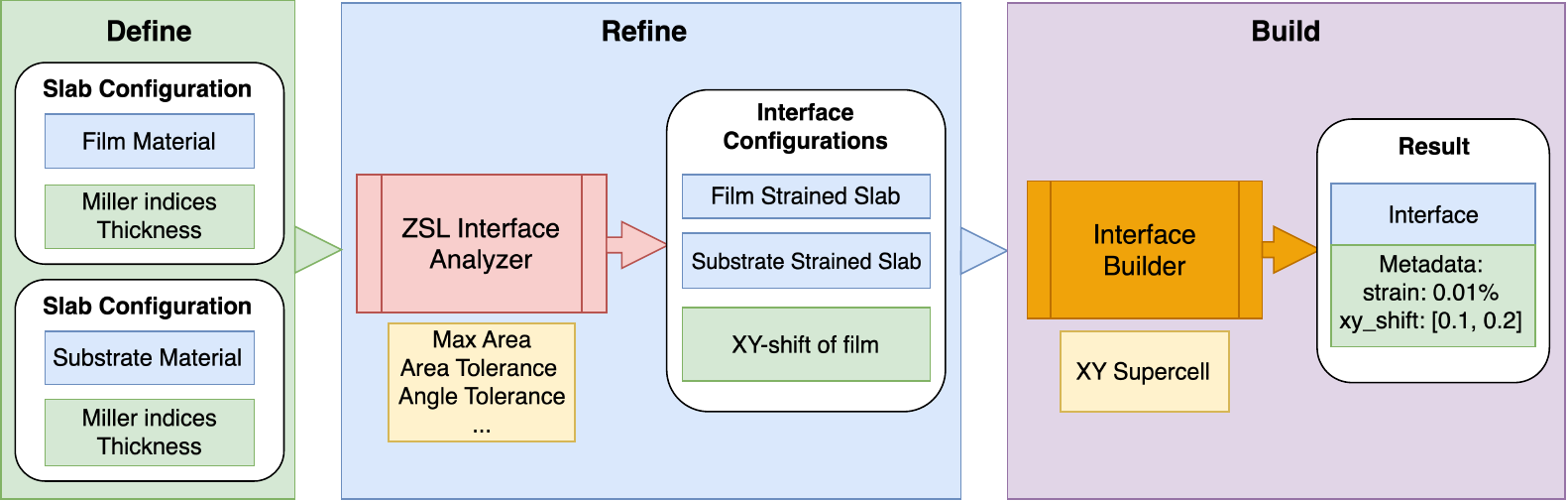}
                \caption{Interface transformation stages: define, refine, and build for selecting and constructing a strain-matched interface with recorded metadata.}
                \label{fig:define-refine-build}
            \end{figure}

            A key pattern in this layer is provenance-aware transformation: materials are created or modified through explicit configurations and recorded build steps. Some transformations can be expressed as a single build step, while others require staged workflows. For example, strain-matched interface generation can be expressed as \textit{Define--Refine--Build}: define film and substrate slabs, refine by enumerating and ranking commensurate matches, and build the selected interface while recording strain and relative shifts as metadata. Figure~\ref{fig:define-refine-build} summarizes this staged workflow, and Section~\ref{sec:results} provides a concrete Gr/Ni(001) example with API snippets.

        \subsubsection{Notebook examples and browser access}
        \label{subsubsec:methodology-components-examples}

            The user-facing examples are distributed through \texttt{mat3ra-api-examples}~\cite{mat3ra-api-examples-pypi-package}. This layer shows how the stack is used in practice through reusable Jupyter notebooks. The collection includes generic notebooks for individual transformations and multi-step workflow notebooks that combine several operations into more advanced examples. These notebooks serve both as documentation and as reproducible templates for adaptation.

            The notebook layer is designed to operate in conventional Python environments and in browser-based runtimes. In particular, the same examples can be executed through JupyterLite using Pyodide, enabling inspection and execution without local installation. This browser-accessible layer is important for accessibility and dissemination, while the Results section shows the kinds of structures and notebook collections that can be built on top of it.

    \subsection{Open-source ecosystem approach}
    \label{subsec:methodology-ecosystem}

        Taken together, these layers form an open-source ecosystem rather than a single application. Standards, common data, implementation packages, and notebooks are distributed separately so that they can be reused independently or together, depending on the use case. This separation of concerns supports validation, integration into other software systems, and gradual extension of the platform.

        We therefore view Mat3ra-2D not only as a code package, but as a reusable stack for realistic structure generation with provenance. We welcome community contributions that extend the schemas, reference data, transformation tooling, and notebook examples, and that broaden the coverage of realistic structure categories relevant to modeling and AI/ML applications.

    \section{Results}
\label{sec:results}

    \subsection{Build procedures}
    
        Mat3ra-2D enables systematic generation of structures that are central to applications but underrepresented in bulk-centric datasets. The images in Table~\ref{tab:generic-notebooks-table} illustrate a number of representative structures generated using the framework, spanning interfaces, defects, reconstructions, and heterostructures. The build procedures implement typical design tasks via helper functions inside $mat3ra-made$\cite{mat3ra-made-pypi-package}. We demonstrate the approach below.
        
        \subsubsection{Example 1: constructing slabs with specific terminations}
        \newcommand{\TriPanelFigSrTiO}[1]{\makebox[\linewidth][c]{\includegraphics[height=5cm,keepaspectratio]{#1}}}
        \begin{figure}[H]
            \centering
            {\setlength{\tabcolsep}{3pt}%
            \begin{tabularx}{\linewidth}{@{}>{\centering\arraybackslash}X>{\centering\arraybackslash}X>{\centering\arraybackslash}X@{}}
            \TriPanelFigSrTiO{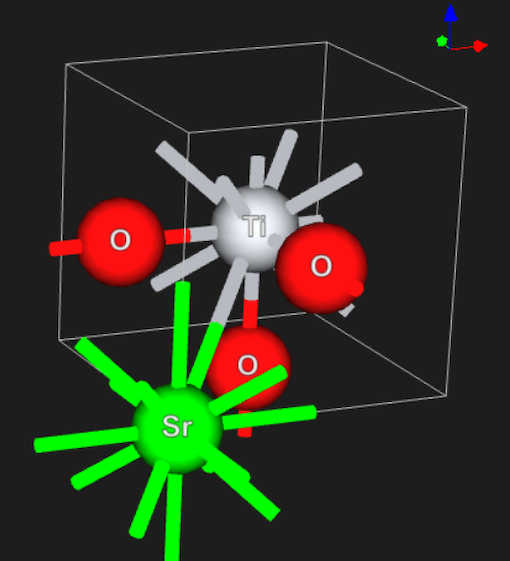} &
            \TriPanelFigSrTiO{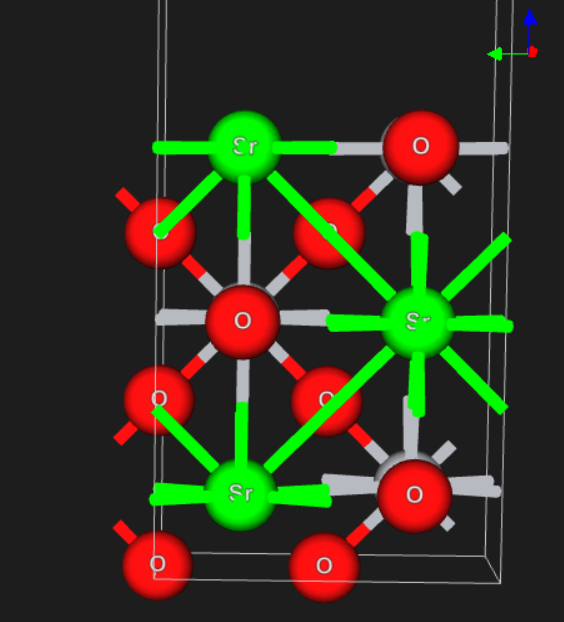} &
            \TriPanelFigSrTiO{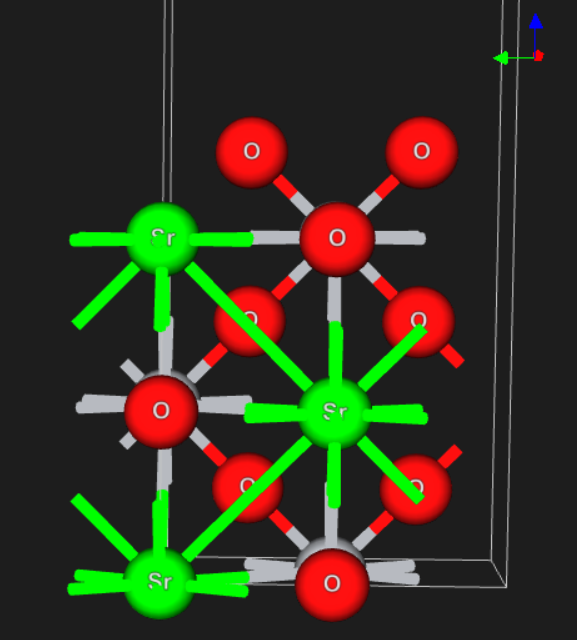} \\
            \small (a) Bulk SrTiO$_3$ & \small (b) (110) slab, SrTiO termination & \small (c) (110) slab, O$_2$ termination
            \end{tabularx}}
            \caption{SrTiO$_3$(110) slab construction with termination control: (a) bulk crystal from the reference data repository, (b) and (c) alternative surface terminations obtained with the same Miller indices, layer count, and vacuum by changing the termination formula.}
            \label{fig:srtio3-slab-terminations}
        \end{figure}
    
        Generation of orientation-specific slabs with explicit termination control is straightforward: for example, a SrTiO$_3$(110) slab is created by importing bulk SrTiO$_3$ from the reference data repository, then applying a slab builder with Miller indices (1,1,0), specifying the number of layers, vacuum spacing, and the desired termination formula (e.g., "SrTiO" or "O$_2$"). Figure~\ref{fig:srtio3-slab-terminations} illustrates the bulk input and the two terminations. This enables systematic exploration of surface chemistry by generating alternative terminations within the same workflow and comparing their resulting structures and properties.
    
\begin{lstlisting}[language=Python]
from mat3ra.standata.materials import Materials
from mat3ra.made.material import Material
from mat3ra.made.tools.build.slab import create_slab

bulk_srtio3 = Material(Materials.get_by_name_first_match("SrTiO3"))
srtio3_slab = create_slab(
    crystal=bulk_srtio3,
    miller_indices=(1, 1, 0),
    termination_formula="SrTiO",
    number_of_layers=4,
    vacuum=10.0,
)
\end{lstlisting}
    
    \subsubsection{Example 2: constructing interfaces without strain matching}
    
        \newcommand{\TriPanelFigGeSi}[1]{\makebox[\linewidth][c]{\includegraphics[height=8.5cm,keepaspectratio]{#1}}}
        \begin{figure}[t]
            \centering
            {\setlength{\tabcolsep}{3pt}%
            \begin{tabularx}{\linewidth}{@{}>{\centering\arraybackslash}X>{\centering\arraybackslash}X>{\centering\arraybackslash}X@{}}
            \TriPanelFigGeSi{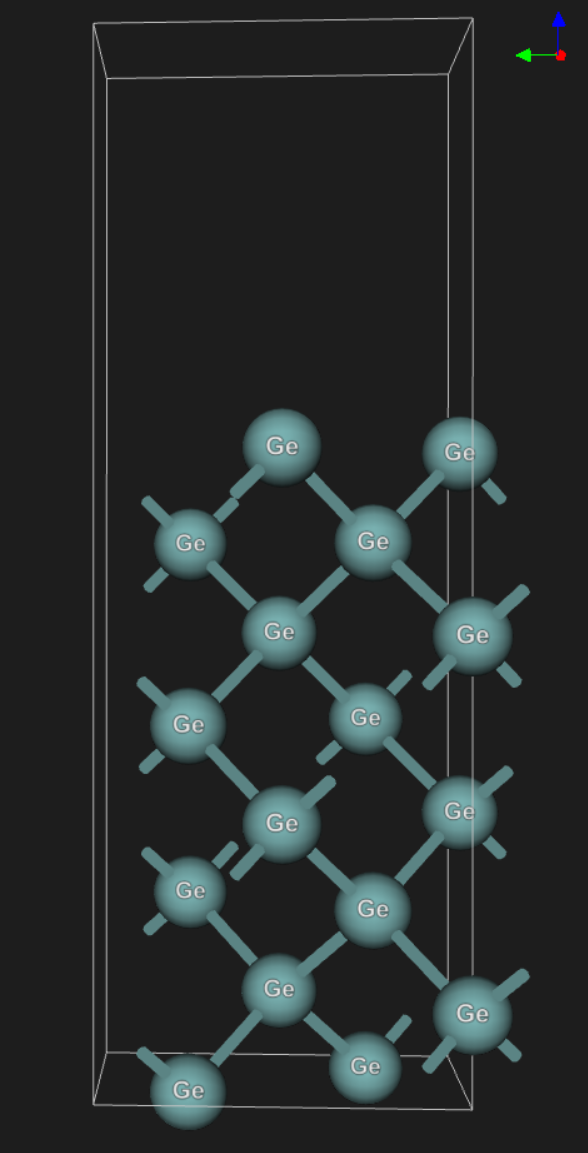} &
            \TriPanelFigGeSi{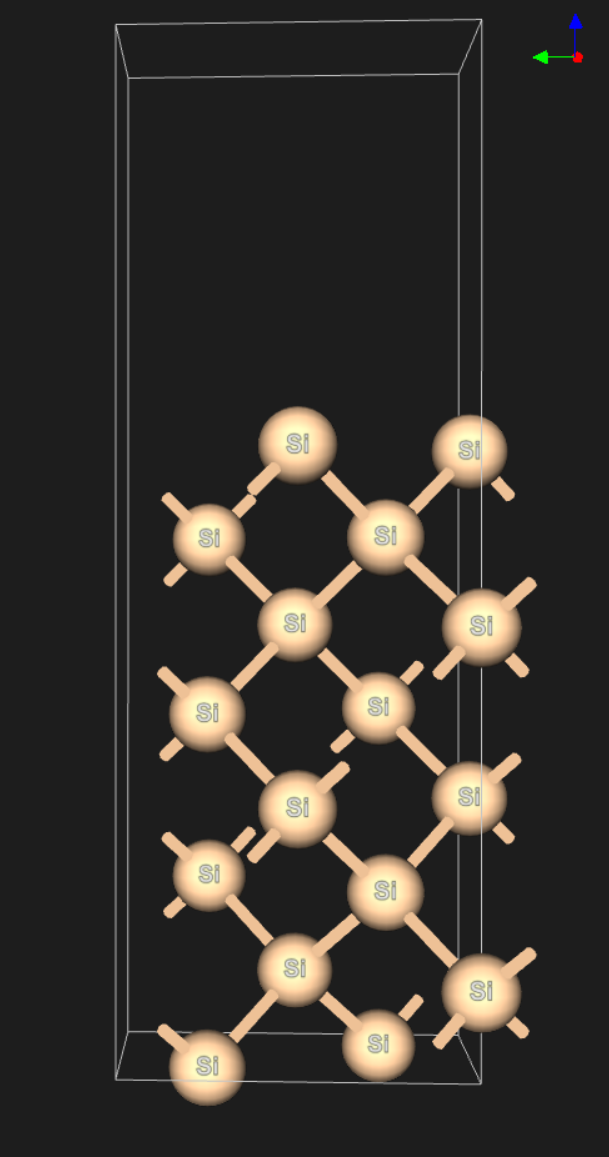} &
            \TriPanelFigGeSi{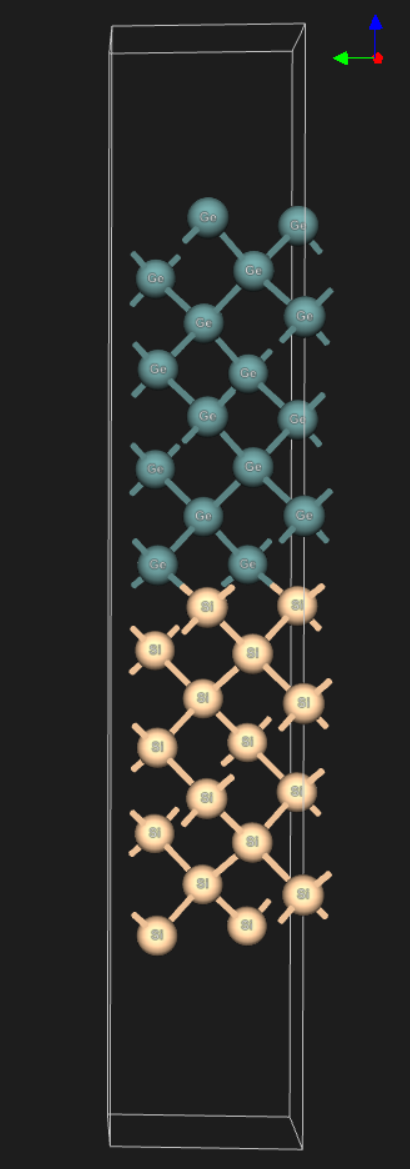} \\
            \small (a) Ge(001) slab & \small (b) Si(001) slab & \small (c) Ge/Si(001) interface
            \end{tabularx}}
            \caption{Ge(001)/Si(001) interface without strain matching: (a) and (b) film and substrate slabs defined independently with Miller indices (001), layer count, and vacuum; (c) combined interface after applying a target interfacial gap and optional lateral shift.}
            \label{fig:ge-si-interface-simple}
        \end{figure}
    
        Assembly of film/substrate interfaces without strain matching follows a define--build workflow. For example, a Ge(001)/Si(001) interface is constructed by defining both slabs independently (each with Miller indices (001), thickness, and vacuum), then combining them with a target interfacial gap (e.g., 1.2~\AA) and optional lateral shift. Figure~\ref{fig:ge-si-interface-simple} illustrates the two slabs and the assembled interface. The resulting interface structure includes metadata recording the source materials and transformation parameters.
    
\begin{lstlisting}[language=Python]
from mat3ra.standata.materials import Materials
from mat3ra.made.material import Material
from mat3ra.made.tools.build.slab import create_slab
from mat3ra.made.tools.build.interface import create_interface_simple_between_slabs

bulk_ge = Material(Materials.get_by_name_first_match("Ge"))
bulk_si = Material(Materials.get_by_name_first_match("Si"))

film_slab = create_slab(
    crystal=bulk_ge,
    miller_indices=(0, 0, 1),
    termination_formula="Ge",
    number_of_layers=2,
)

substrate_slab = create_slab(
    crystal=bulk_si,
    miller_indices=(0, 0, 1),
    termination_formula="Si",
    number_of_layers=2,
)

interface = create_interface_simple_between_slabs(
    substrate_slab=substrate_slab,
    film_slab=film_slab,
    gap=1.2,
)
\end{lstlisting}
    
    \subsubsection{Example 3: strain-matched interfaces}

\begin{figure}[H]
            \centering
            \begin{tabular}{cc}
            \includegraphics[width=0.48\textwidth]{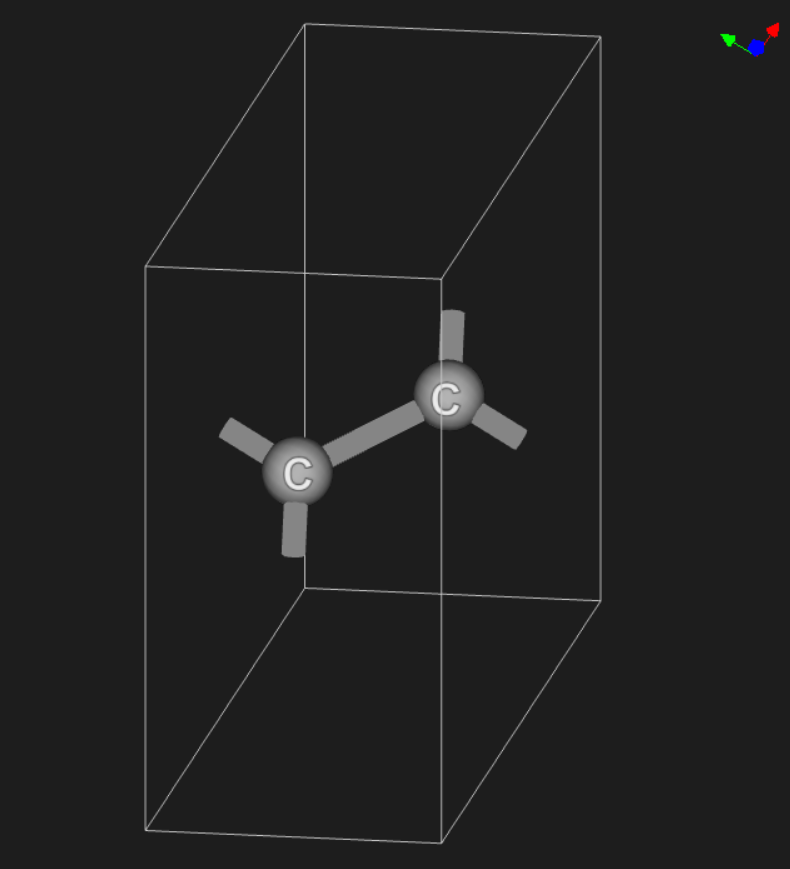} &
            \includegraphics[width=0.48\textwidth]{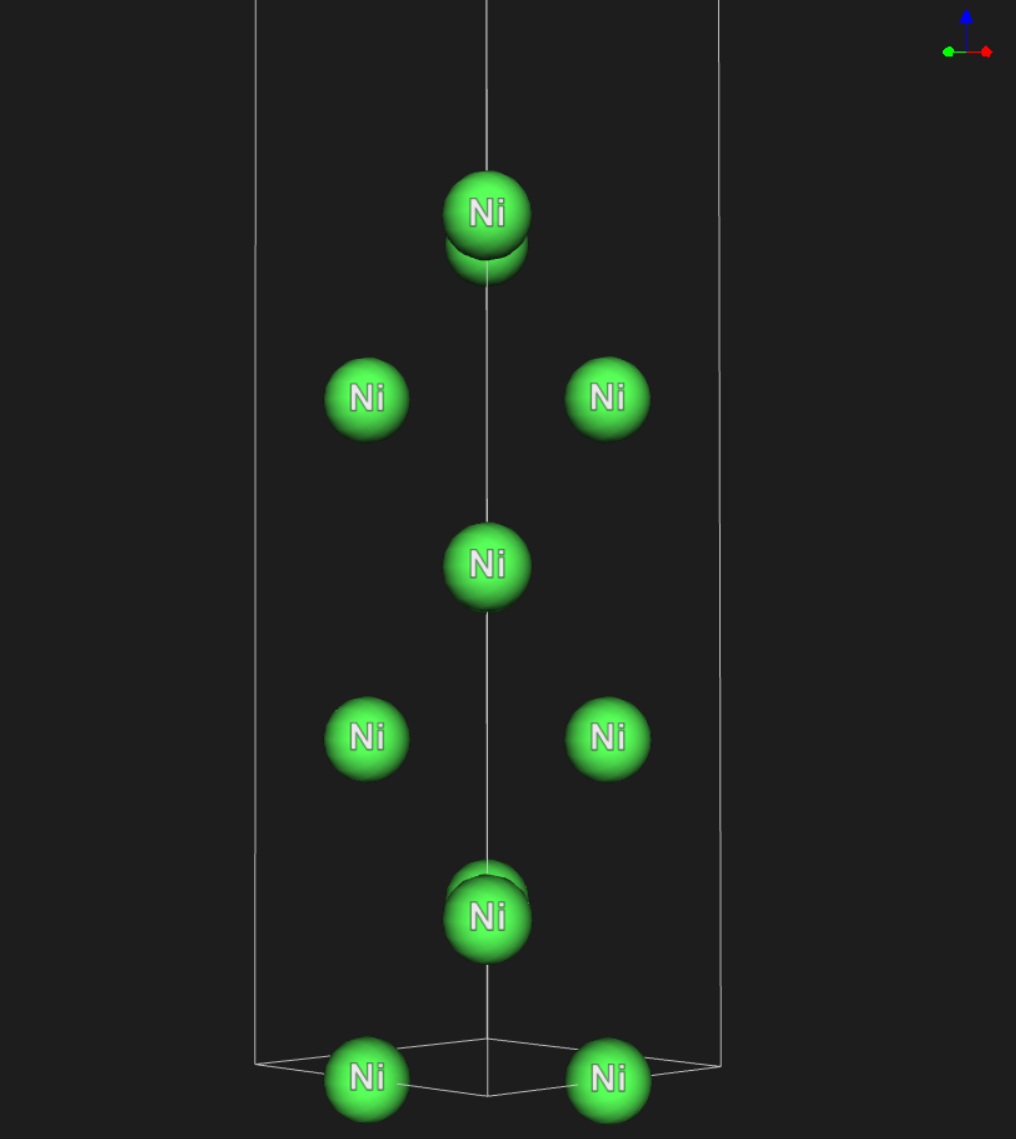} \\
            (a) Graphene monolayer & (b) Ni(001) slab \\
            \includegraphics[width=0.48\textwidth]{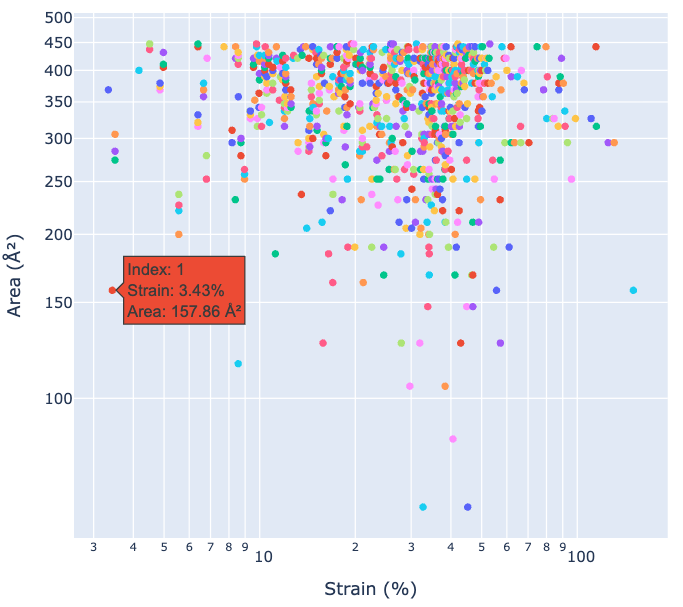} &
            \includegraphics[width=0.48\textwidth]{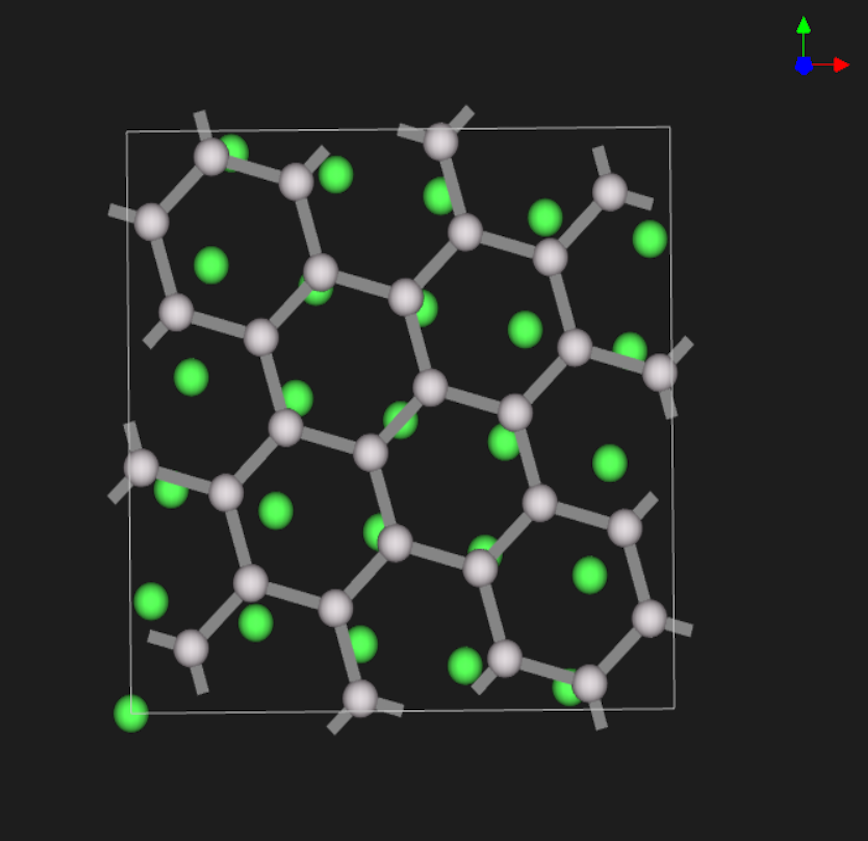} \\
            (c) Strain--size trade-off & (d) Selected Gr/Ni(001) interface
            \end{tabular}
            \caption{Graphene/Ni(001) interface construction workflow: (a) Starting graphene monolayer film, (b) Ni(001) slab substrate, (c) ZSL analyzer output showing strain--size trade-off for candidate configurations (each point represents a commensurate match ranked by strain percentage and interface area), (d) Selected interface structure with commensurate matching applied. The workflow follows define--refine--build: define film and substrate slabs, refine by enumerating and ranking matches, then build the selected configuration with recorded metadata.}
            \label{fig:gr-ni-workflow}
        \end{figure}

        For interfaces requiring commensurate matching, the framework provides a define--refine--build workflow. Figure~\ref{fig:gr-ni-workflow} illustrates this process for constructing a Graphene/Ni(001) interface.
        
        \underline{Define}: Starting from bulk materials (Figures~\ref{fig:gr-ni-workflow}a--b), the film (graphene monolayer) and substrate (Ni(001) slab) configurations are created independently. Bulk crystals are imported from reference data and wrapped as materials, after which a slab builder is applied using Miller indices, layer counts, vacuum spacing, and termination formulas. The graphene film is a single-layer structure with (001) orientation, while the Ni substrate is a (001) slab with multiple layers.
        
        \underline{Refine}: A ZSL (Zur and McGill superlattice) analyzer enumerates commensurate matches by finding supercell combinations that minimize strain while controlling interface area. Figure~\ref{fig:gr-ni-workflow}c shows the strain--size trade-off plot, where each point represents a candidate configuration ranked by strain percentage and number of atoms (interface area). The analyzer returns multiple ranked options, allowing selection based on the desired balance between strain minimization and computational cost (system size).
        
        \underline{Build}: The selected configuration is passed to the interface builder, which applies the supercell transformations to the film and substrate slabs, enforces the chosen interfacial gap and lateral shift, adds vacuum if requested, and records strain and relative shift as metadata. Figure~\ref{fig:gr-ni-workflow}d shows the resulting Graphene/Ni(001) interface structure with commensurate matching applied.
                
        In practice, bulk crystals are retrieved from the reference data repository by name (e.g., ``Graphene'' and ``Nickel'') and wrapped as materials for downstream building and analysis (Figure~\ref{fig:gr-ni-workflow}a--b).
    
\begin{lstlisting}[language=Python]
from mat3ra.standata.materials import Materials
from mat3ra.made.material import Material

bulk_graphene = Material(Materials.get_by_name_first_match("Graphene"))
bulk_nickel = Material(Materials.get_by_name_first_match("Nickel"))
\end{lstlisting}

        The film and substrate slabs are then defined independently by applying a slab builder with Miller indices, termination formula, layer count, and vacuum spacing. In this example, graphene is defined as a single layer with C termination, while nickel is defined as a (001) slab with Ni termination and 3 layers.
    
\begin{lstlisting}[language=Python]
from mat3ra.made.tools.build.slab import create_slab

film_slab = create_slab(
    crystal=bulk_graphene,
    miller_indices=(0, 0, 1),
    termination_formula="C",
    number_of_layers=1,
    vacuum=10.0,
)

substrate_slab = create_slab(
    crystal=bulk_nickel,
    miller_indices=(0, 0, 1),
    termination_formula="Ni",
    number_of_layers=3,
    vacuum=10.0,
)
\end{lstlisting}
    
        To refine the interface, a ZSL analyzer enumerates commensurate supercell matches subject to a maximum area constraint and returns a ranked list of candidate configurations (Figure~\ref{fig:gr-ni-workflow}c). The top-ranked candidate provides a strained film configuration paired with a matching substrate configuration.
    
\begin{lstlisting}[language=Python]
from mat3ra.made.tools.build.interface import ZSLInterfaceAnalyzer

analyzer = ZSLInterfaceAnalyzer(
    film=film_slab,
    substrate=substrate_slab,
    max_area=50.0,
)
configurations = analyzer.get_configurations()
strained_film, substrate = configurations[0]
\end{lstlisting}
    
        Finally, the selected configurations are passed to an interface builder that constructs the combined interface structure while enforcing an interfacial gap, vacuum spacing, and an optional lateral shift for the film (Figure~\ref{fig:gr-ni-workflow}d).
    
\begin{lstlisting}[language=Python]
from mat3ra.made.tools.build.interface import create_interface

interface = create_interface(
    film_configuration=strained_film,
    substrate_configuration=substrate,
    gap=2.1,
    vacuum=10.0,
    xy_film_shift=[0.0, 0.0],
)
\end{lstlisting}
        
        This workflow enables systematic exploration of interface configurations while maintaining provenance of the matching process, enabling reproducible dataset generation and transparent reporting of generation parameters.

    \subsection{Jupyter notebooks demonstrating the usage}
    
    The structural build procedures are demonstrated and shared as reusable Jupyter notebooks that serve as both interactive documentation and reproducible templates. The notebooks are written in an interactive and editable way that accepts multiple input parameters, summarized at the top, and editable by users. The inputs include the initial materials structures (e.g. - film and substrate for interface creation), and the parameters of the builder and subsequent downselection. These notebooks demonstrate how to use individual transformations (e.g., slab creation, interface construction, passivation). These notebooks are designed to be educational and adaptable: they include functionality for loading materials from the reference data repository, previewing structures, saving results, and provide default settings and parameters. Users can run these notebooks to understand the transformation approach, then adjust parameter values to create their desired structures. Each notebook focuses on a single transformation type, making it easy to learn and modify. The notebooks are available online at \href{https://jupyterlite.mat3ra.com}{jupyterlite.mat3ra.com} through a browser-based JupyterLite environment, making the examples directly inspectable and executable in any modern web browser without local installation or environment setup.
        Each notebook is annotated with an M-CODE tag \cite{biryukov2026mcode} indicating the corresponding structure category.
        The list of notebooks is presented in Table~\ref{tab:generic-notebooks-table}.

        \begin{figure}[H]
            \centering
            \includegraphics[width=\textwidth]{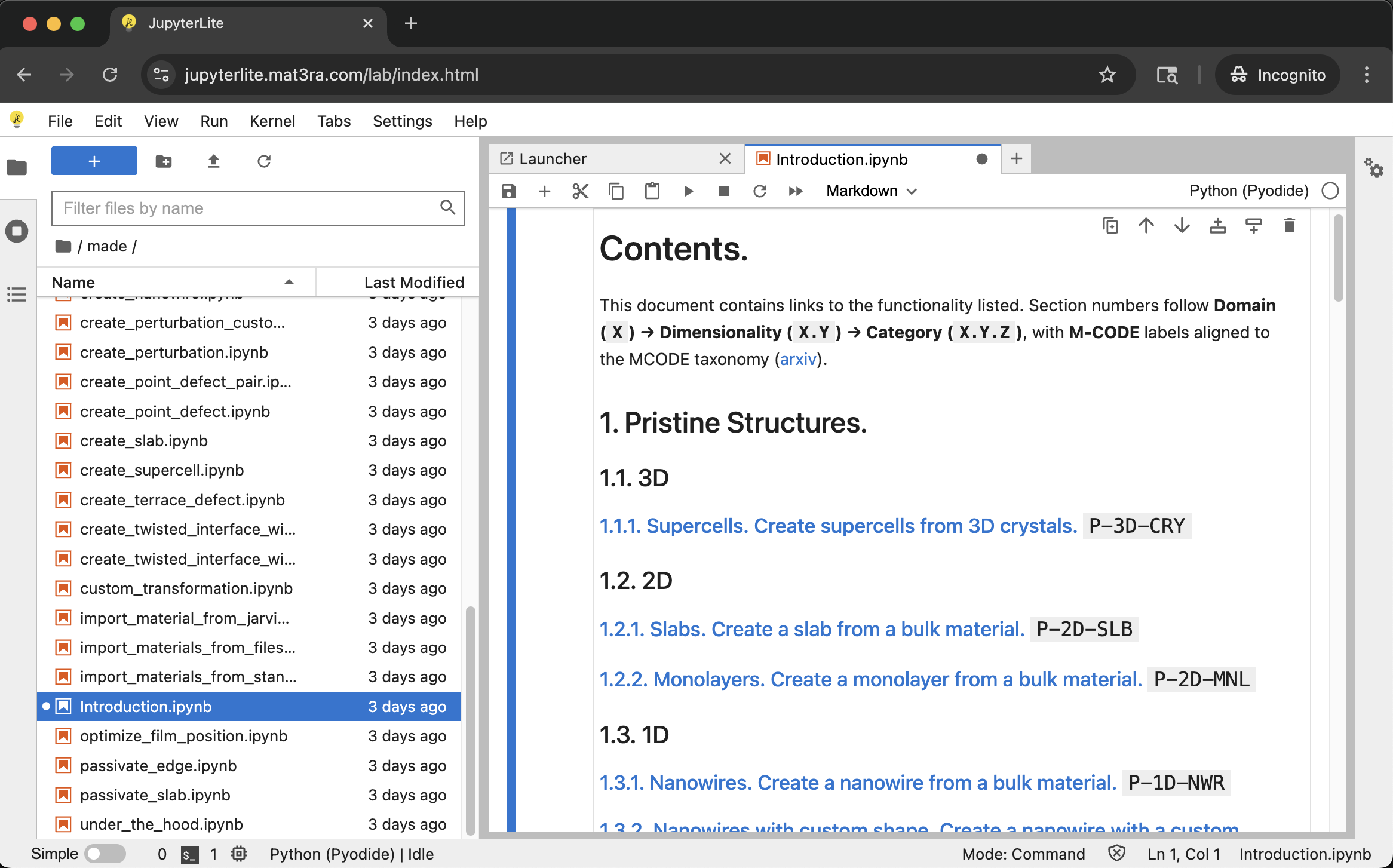}
            \caption{The JupyterLite environment hosting the transformation notebooks. The left panel shows the file system browser with the notebooks listed, the right panel shows the editor with the Introduction notebook open. The Introduction notebook contains the index and allows to navigate to the other ones. The notebooks are organized by the M-CODE\cite{biryukov2026mcode} tag for quick navigation.}
            \label{fig:nb-intro}
        \end{figure}

        \begin{figure}[H]
            \centering
            \includegraphics[width=\textwidth]{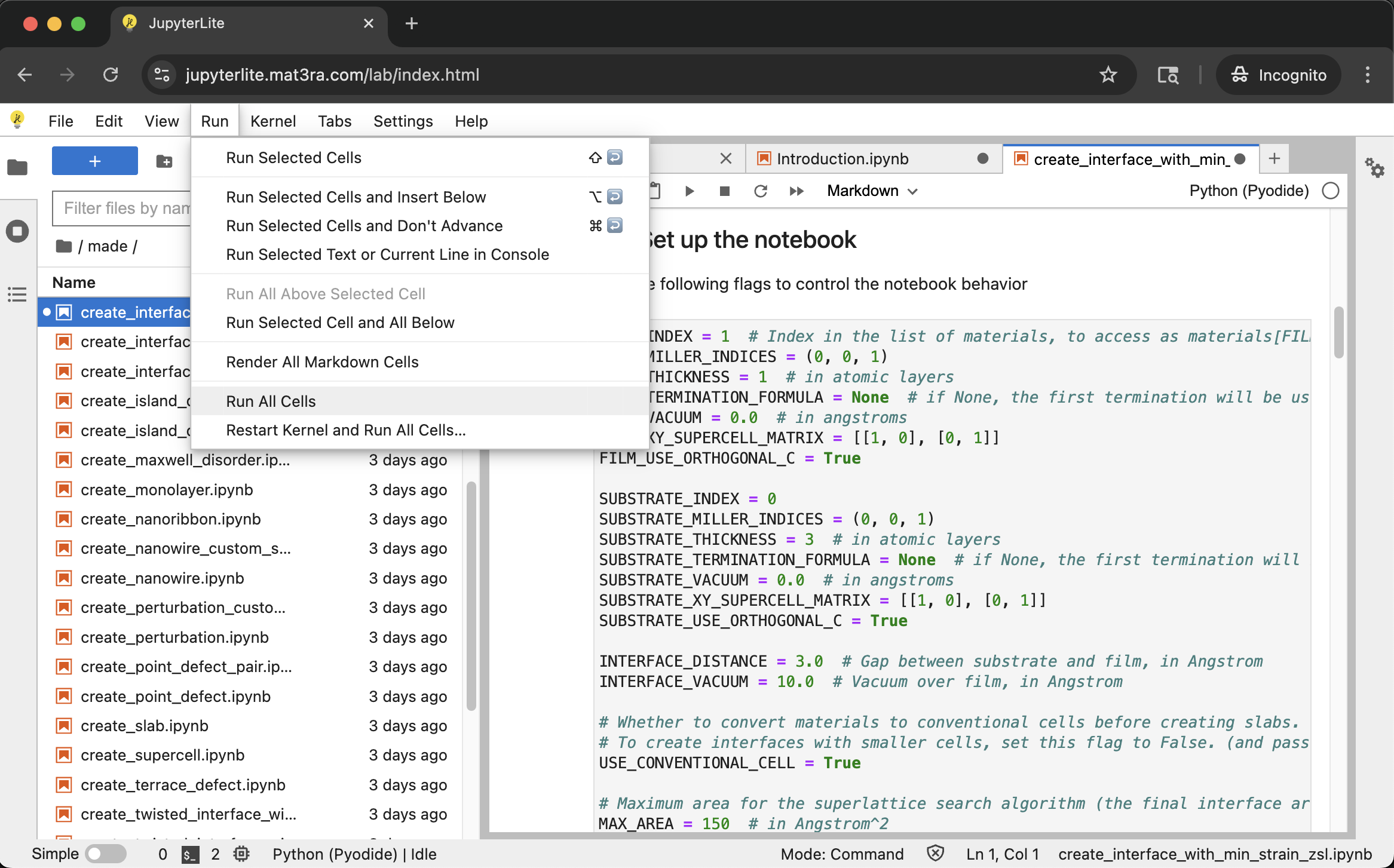}
            \caption{Example material transformation notebook opened in JupyterLite. The notebook exposes editable input cells and executable workflow steps so that users can modify parameters, run the build procedure (via \textit{Run} \texttt{\textgreater} \textit{Run All Cells} menu item), and inspect intermediate results directly in the browser. The example shows the interface construction workflow. The input parameters contain the film and substrate slab configurations, the interfacial gap, and the lateral shift. The settings for the strain matching algorithm - e.g. the maximum area constraint - are also shown.}
            \label{fig:nb-run}
        \end{figure}

        \begin{figure}[H]
            \centering
            \includegraphics[width=\textwidth]{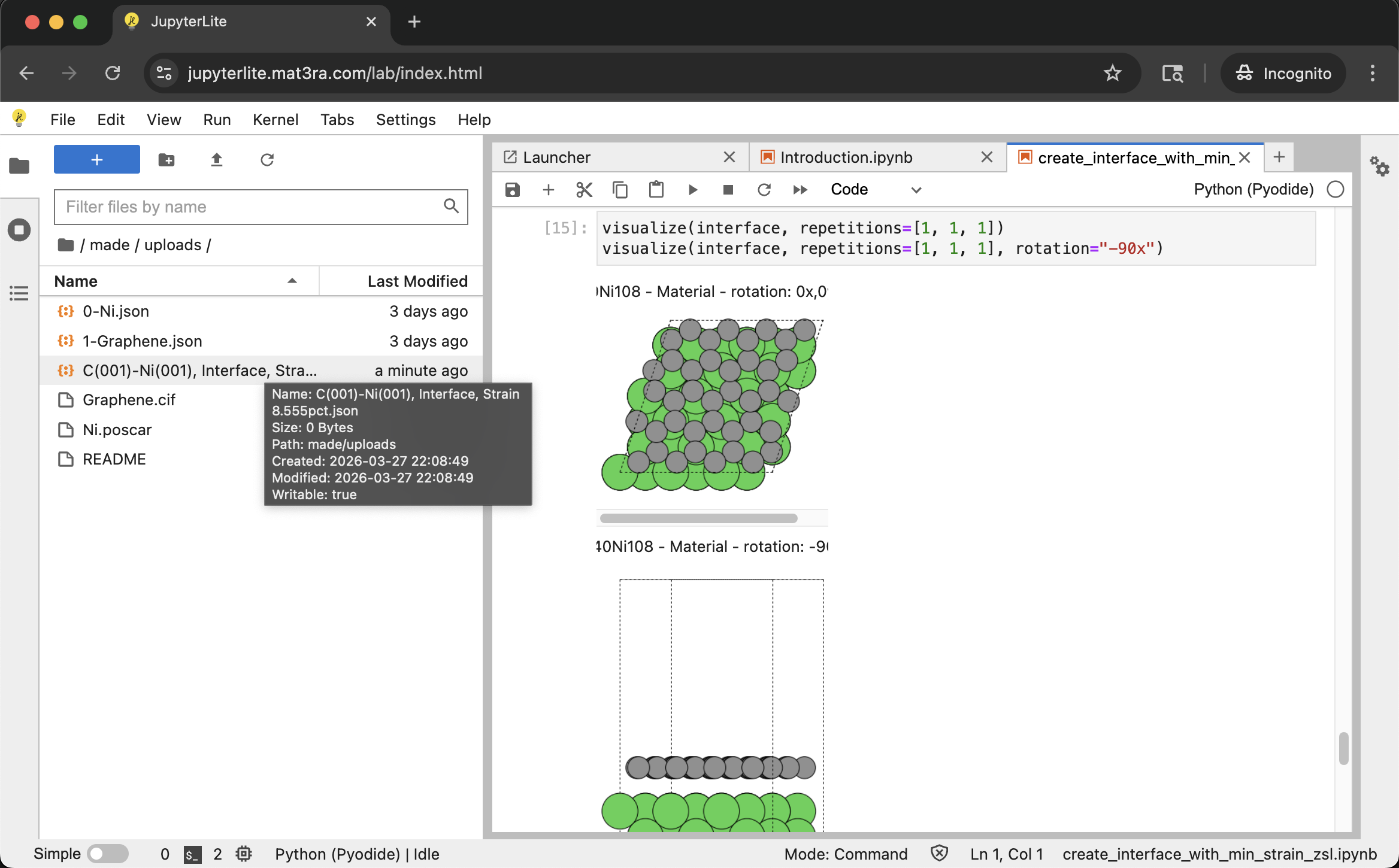}
            \caption{Final output produced by an example notebook in JupyterLite. The resulting structure can be previewed after execution, illustrating how the notebooks serve as interactive and reproducible templates for realistic structure generation. The example shows the final result of the Gr/Ni(001) interface construction workflow. The final structure is saved in the \textit{uploads} folder and can be loaded back into the (other) notebook(s) environment for further analysis or reuse.}
            \label{fig:nb-final-result}
        \end{figure}

        \newpage
        \setlength{\LTleft}{0pt}
        \setlength{\LTright}{0pt}
        \setlength{\LTcapwidth}{\textwidth}
        \begingroup
        \setlength{\tabcolsep}{4pt}
        \MateraTableSetup
        \begin{longtable}{|>{\raggedright\arraybackslash}m{0.12\textwidth}|>{\raggedright\arraybackslash}m{0.2\textwidth}|>{\raggedright\arraybackslash}m{0.31\textwidth}|>{\centering\arraybackslash}m{0.27\textwidth}|}
            \caption{The list of notebooks for material transformations available at \href{https://jupyterlite.mat3ra.com}{jupyterlite.mat3ra.com}. The notebooks are organized according to the M-CODE\cite{biryukov2026mcode} tag corresponding to the resulting structures produced. The Name column includes the name of the notebook. Notes explain the (adjustable by user) input parameters and the example structure shown in the last column.}
            \label{tab:generic-notebooks-table} \\
            \hline
            \MateraTH{M-CODE} & \MateraTH{Name} & \MateraTH{Notes} & \MateraTH{Example} \\
            \hline
            \endfirsthead
            \hline
            \MateraTH{M-CODE} & \MateraTH{Name} & \MateraTH{Notes} & \MateraTH{Example} \\
            \hline
            \endhead
            \texttt{P-3D-CRY} & Load Material from Standata & Input: material identifier from the reference repository.\newline Result: material structure.\newline Example: Ni bulk. & \includegraphics[width=\linewidth]{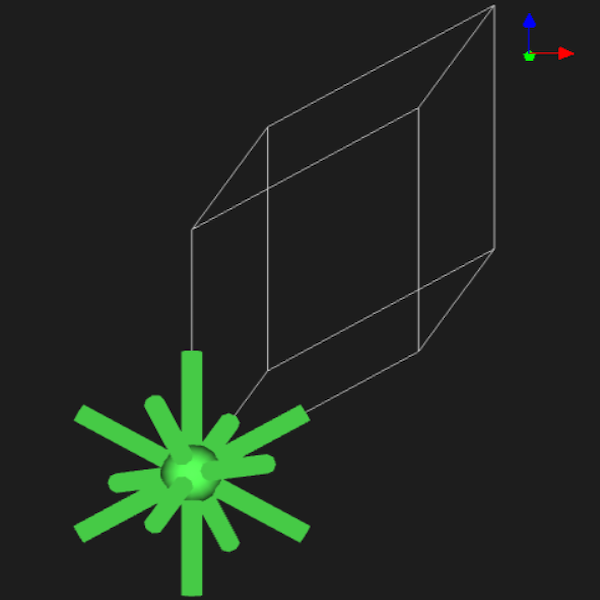} \\ \hline
            \texttt{P-3D-CRY} & Create Supercell & Inputs: bulk crystal and replication matrix.\newline Result: crystal supercell.\newline Example: Si 3x3x3 supercell. & \includegraphics[width=\linewidth]{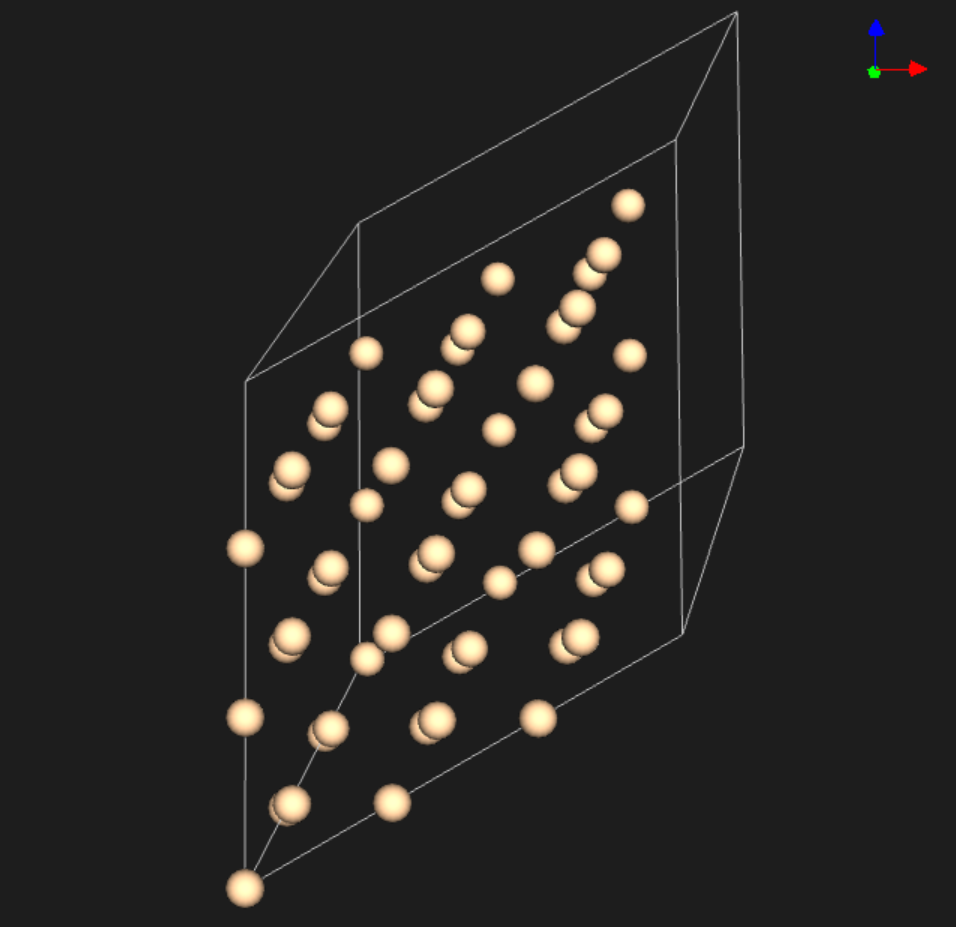} \\ \hline
            \texttt{P-2D-MNL} & Create Monolayer & Input: layered bulk crystal and cleavage settings.\newline Result: isolated 2D monolayer.\newline Example: graphene monolayer. & \includegraphics[width=\linewidth]{data/figures/Graphene.png} \\ \hline
            \texttt{P-2D-SLB-S} & Create Slab & Input: bulk crystal, Miller indices, layers, vacuum, termination.\newline Result: a slab structure.\newline Example: SrTiO$_3$(110) slab with SrTiO termination. & \includegraphics[width=\linewidth]{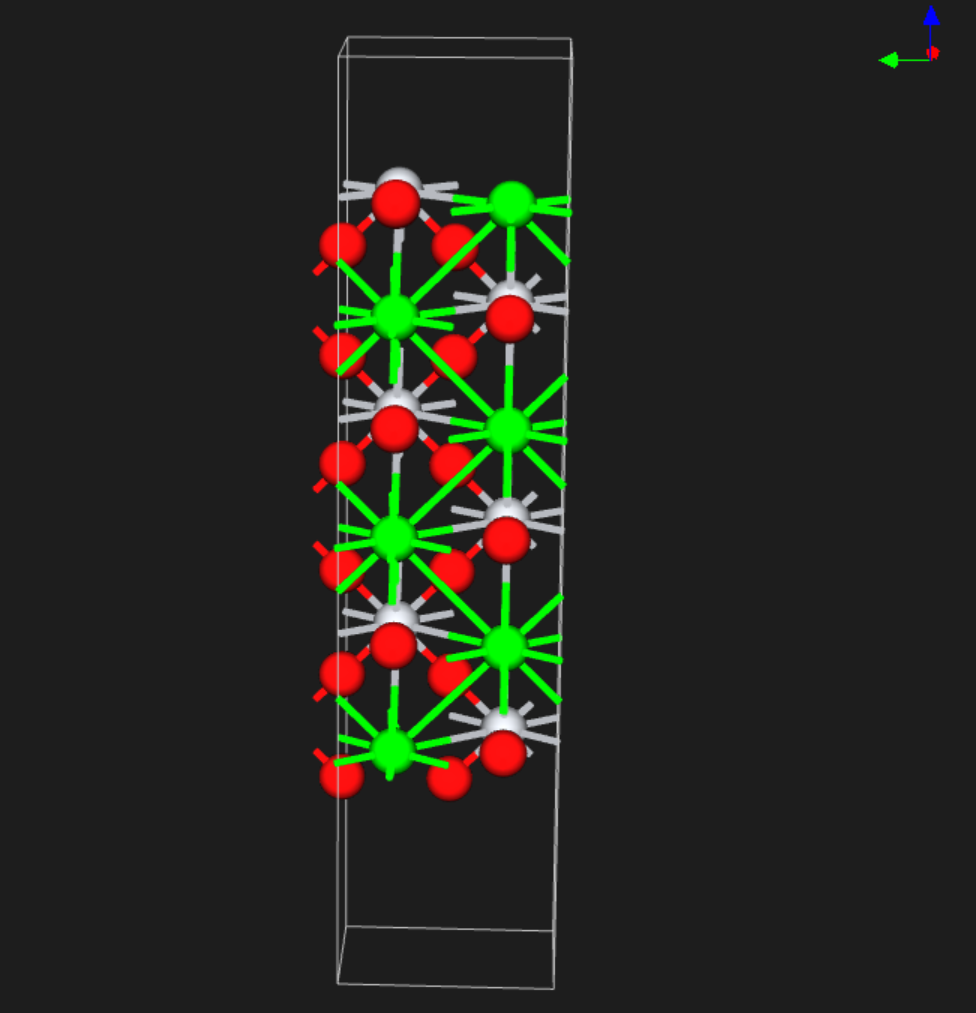} \\ \hline
            \texttt{P-1D-NWR} & Create Nanowire & Input: bulk crystal, orientation, and cross-section settings.\newline Result: finite-width nanowire.\newline Example: Si(001) nanowire. & \includegraphics[width=\linewidth]{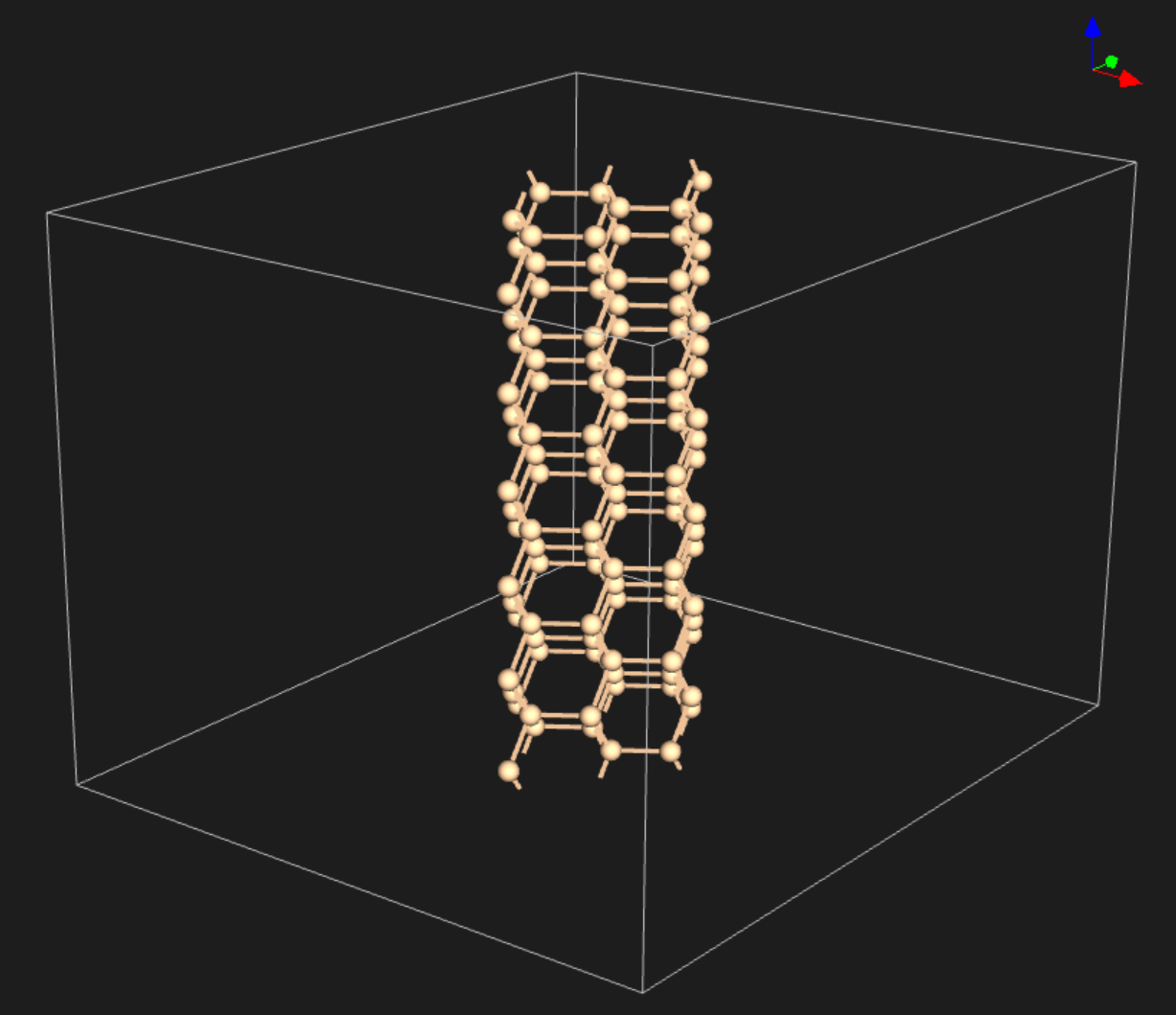} \\ \hline
            \texttt{P-0D-NRB} & Create Nanoribbon & Input: monolayer, cut direction, length, and width.\newline Result: finite nanoribbon.\newline Example: graphene armchair nanoribbon. & \includegraphics[width=\linewidth]{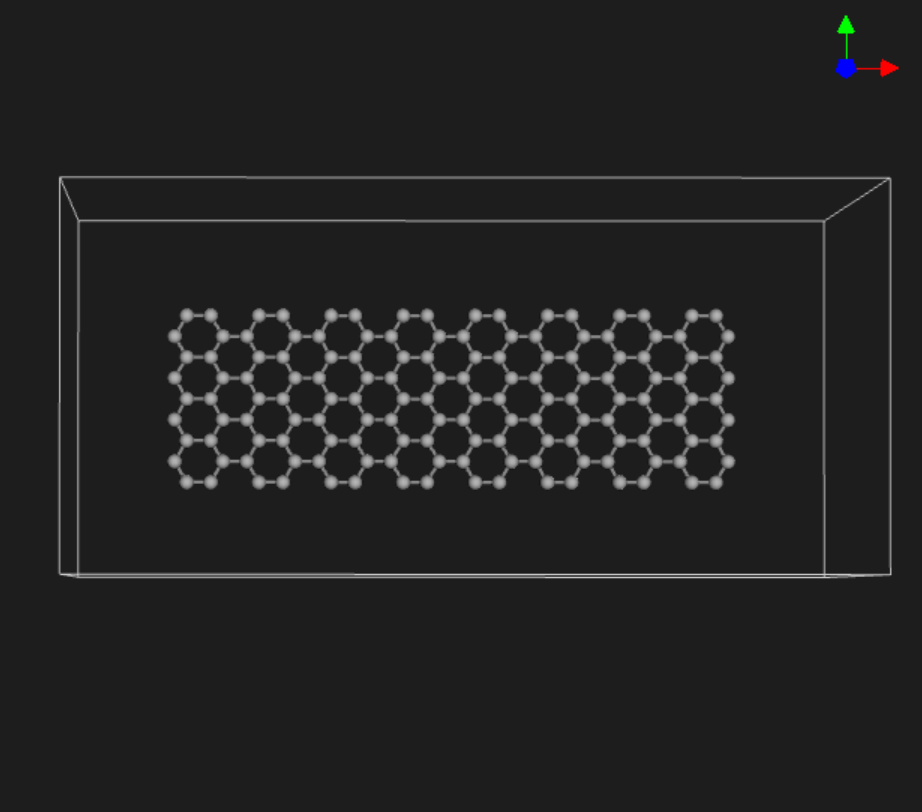} \\ \hline
            \texttt{P-0D-NPR} & Create Cluster & Input: crystal or slab plus shape parameters.\newline Result: finite nanoparticle or cluster.\newline Example: Au nanoparticle. & \includegraphics[width=\linewidth]{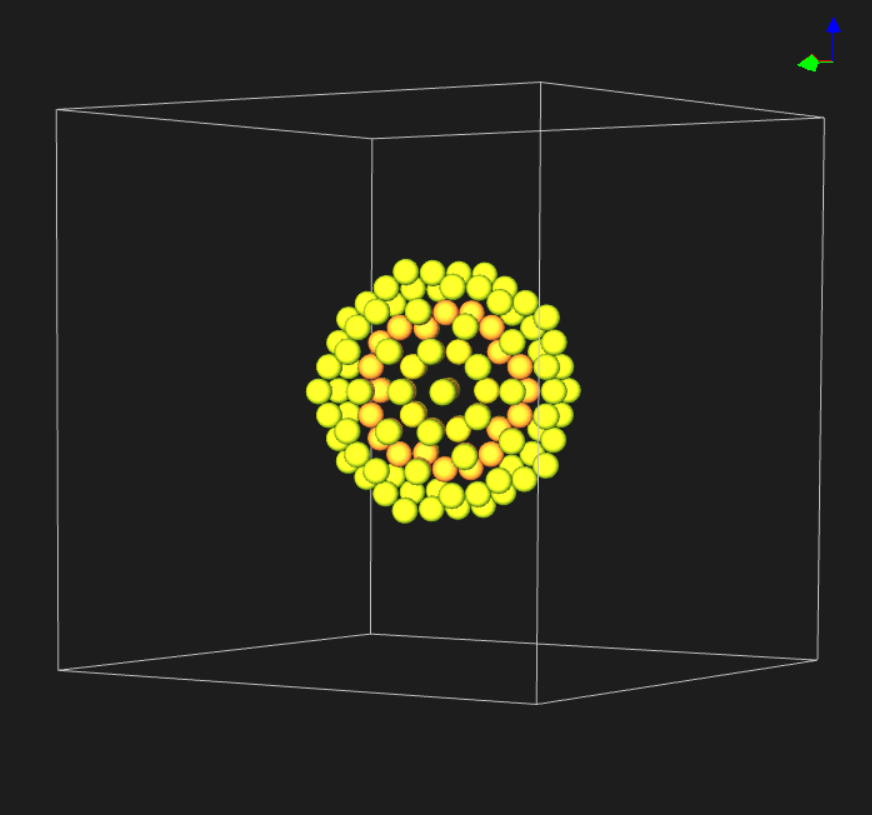} \\ \hline
            \texttt{C-2D-HST} & Create Heterostructure & Input: multiple layers, Miller indices, thickness, and stacking order.\newline Result: layered heterostack.\newline Example: Si/SiO$_2$/HfO$_2$/TiN heterostack. & \includegraphics[width=\linewidth]{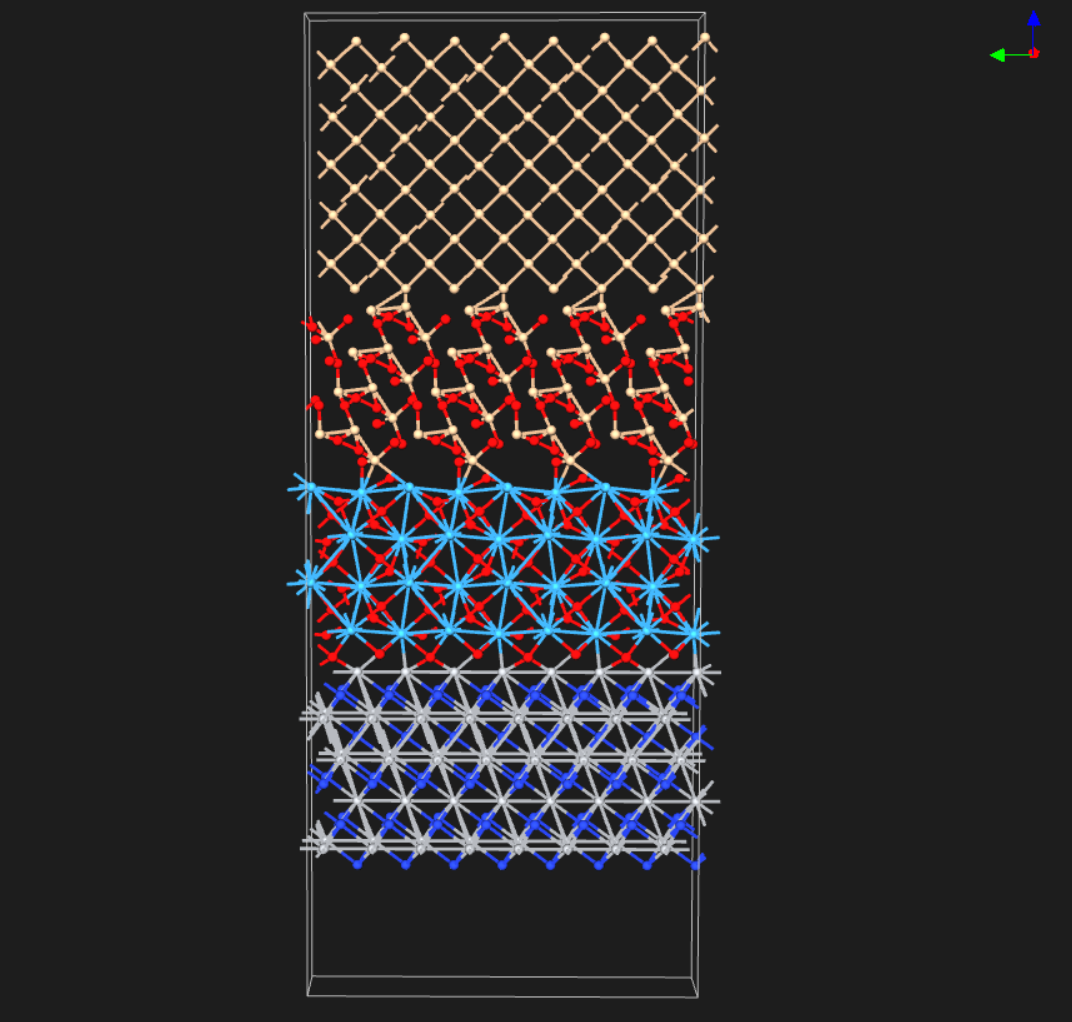} \\ \hline
            \texttt{C-2D-INT-S} & Create Interface with no strain-matching & Input: film and substrate slabs, gap, and shift.\newline Result: directly assembled interface.\newline Example: Ge(001)/Si(001) interface. & \includegraphics[width=\linewidth]{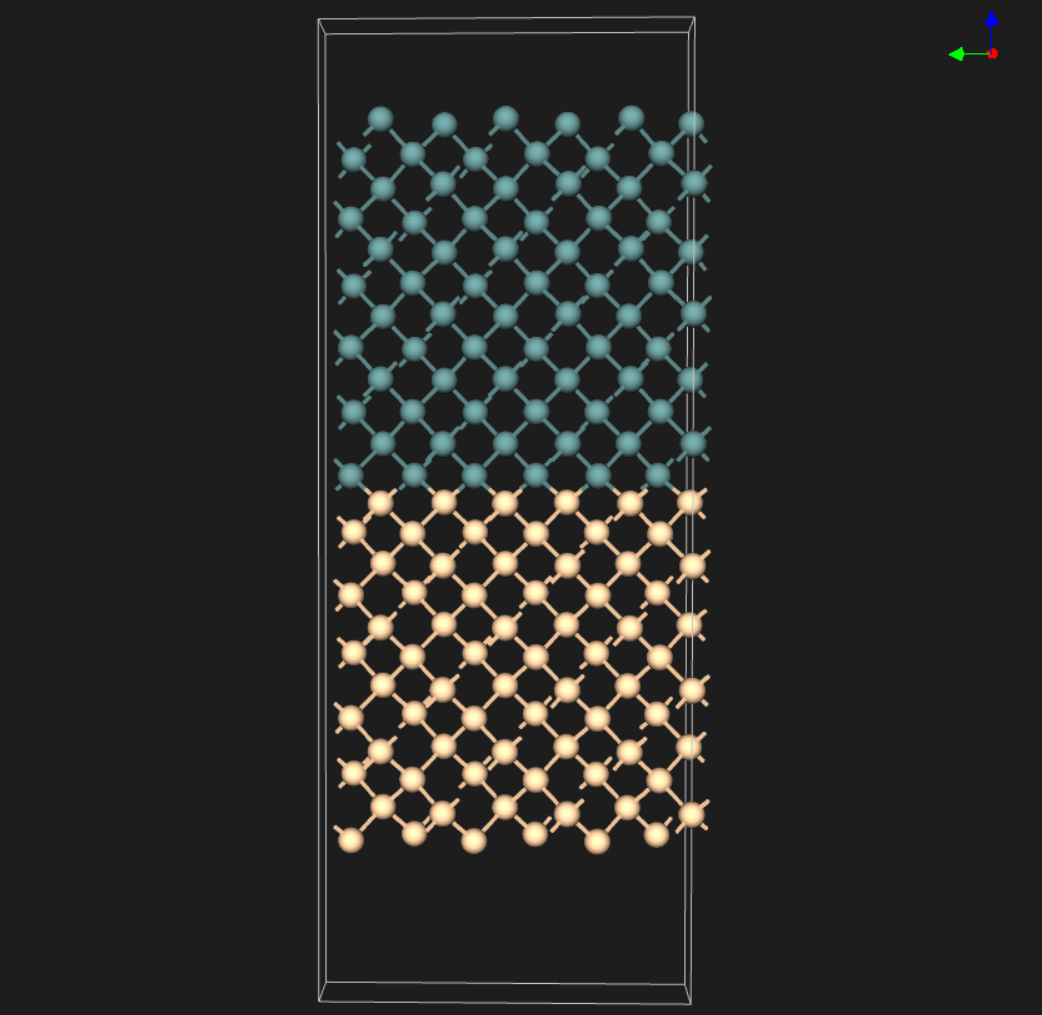} \\ \hline
            \texttt{C-2D-INT-Z} & Create Interface with strain matching ZSL & Input: film and substrate slabs plus ZSL matching constraints.\newline Result: commensurate interface.\newline Example: Graphene/Ni(001) interface. & \includegraphics[width=\linewidth]{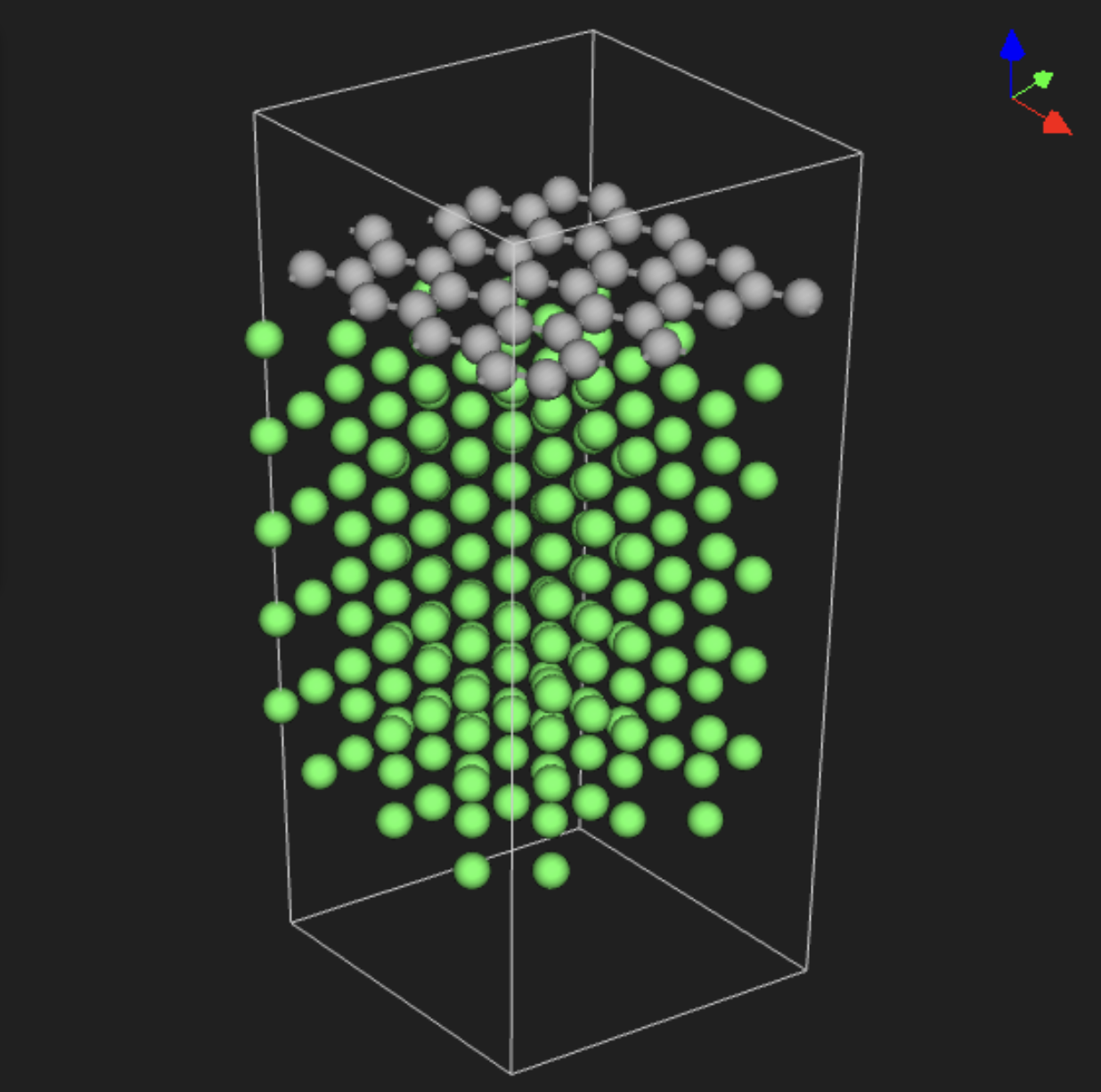} \\ \hline
            \texttt{C-2D-INT-T} & Create twisted interface with commensurate lattices & Input: two layers, twist angle, and matching settings.\newline Result: twisted interface.\newline Example: twisted MoS2/WS2 interface. & \includegraphics[width=\linewidth]{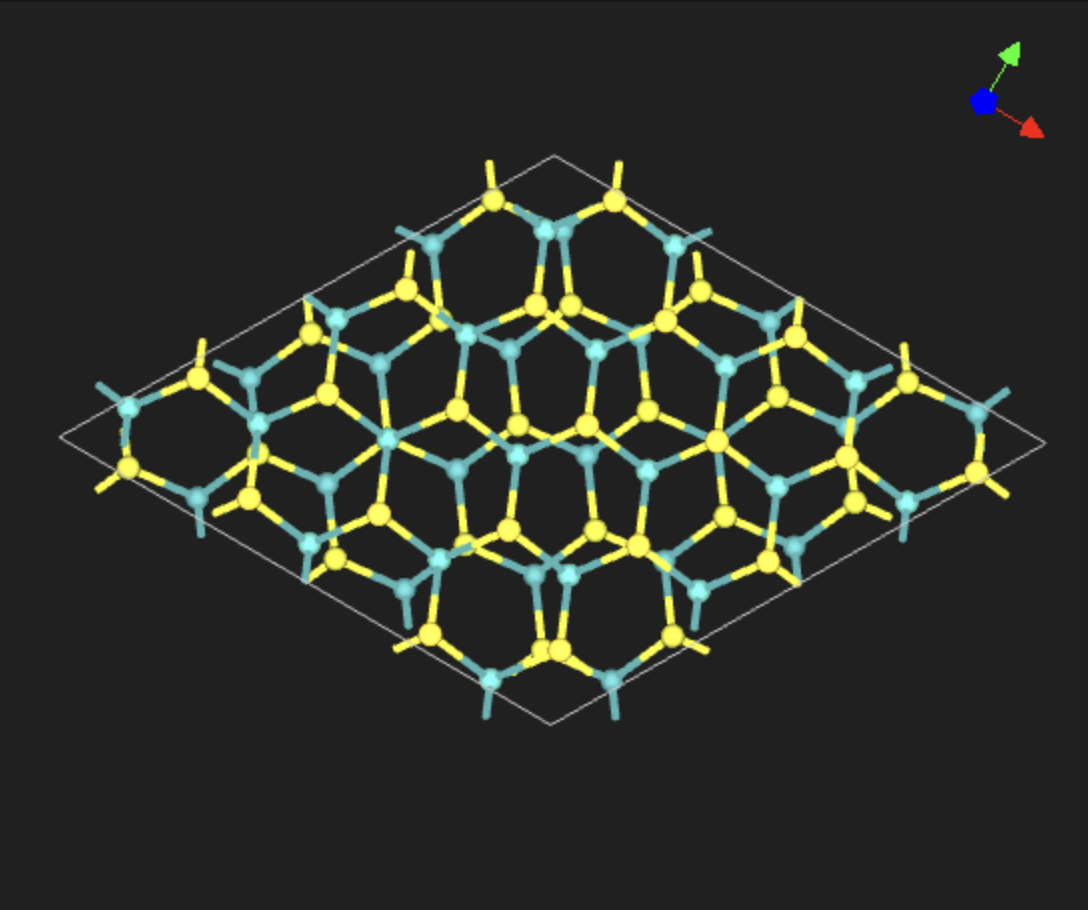} \\ \hline
            \texttt{D-2D-ADA} & Create adatom defect & Input: surface slab, adsorption coordinates, and adatom species.\newline Result: slab with adatom defect.\newline Example: Li adatom on Graphene. & \includegraphics[width=\linewidth]{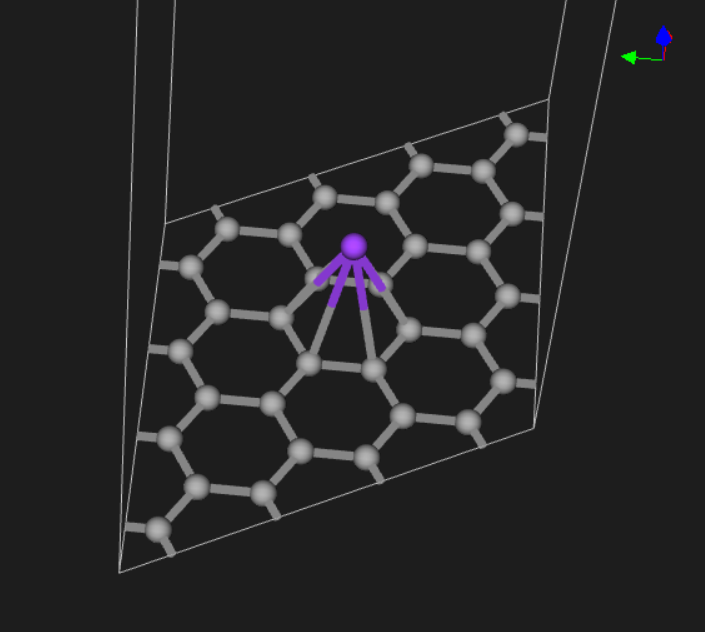} \\ \hline
            \texttt{D-2D-ISL} & Create Island defect & Input: parent surface and island geometry parameters.\newline Result: slab with surface island.\newline Example: TiN island on TiN(001) surface. & \includegraphics[width=\linewidth]{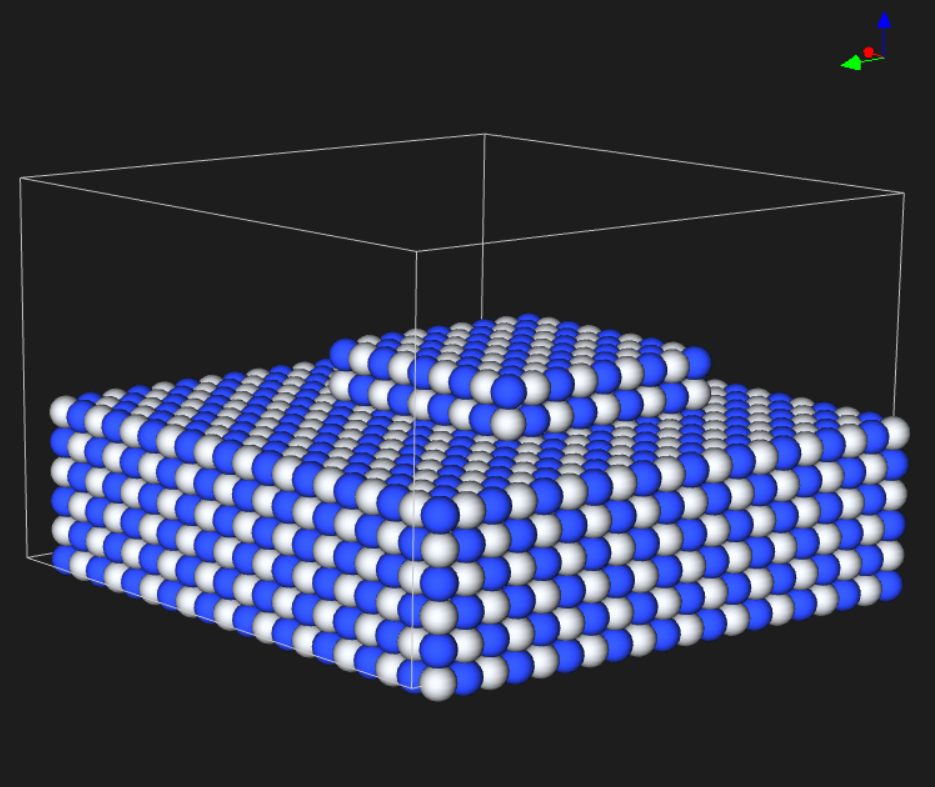} \\ \hline
            \texttt{D-2D-TER} & Create a terrace & Input: slab and terrace cut orientation.\newline Result: stepped surface model.\newline Example: Pt(210) terrace. & \includegraphics[width=\linewidth]{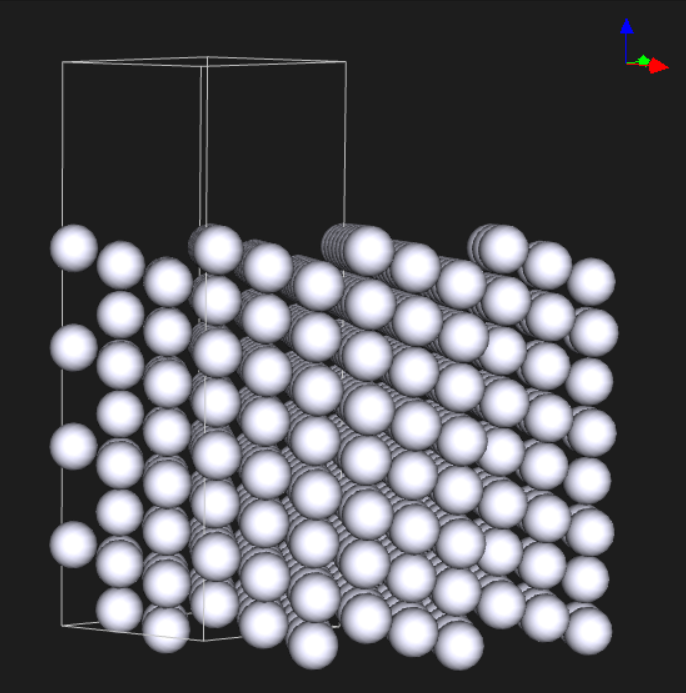} \\ \hline
            \texttt{D-2D-GBP} & Create a grain boundary & Input: two grains and boundary matching parameters.\newline Result: planar grain boundary.\newline Example: Si(001)/(111) grain boundary. & \includegraphics[width=\linewidth]{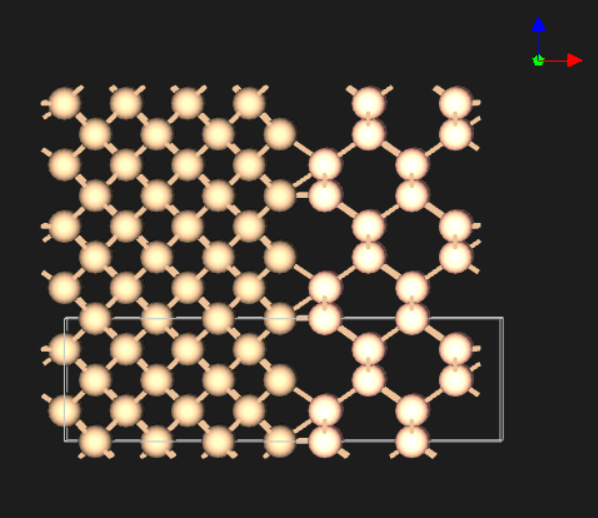} \\ \hline
            \texttt{D-1D-GBL} & Create a Grain Boundary in 2D material & Input: 2D lattice, misorientation, and join settings.\newline Result: line defect in a 2D sheet.\newline Example: grain boundary in hBN at 9 degrees. & \includegraphics[width=\linewidth]{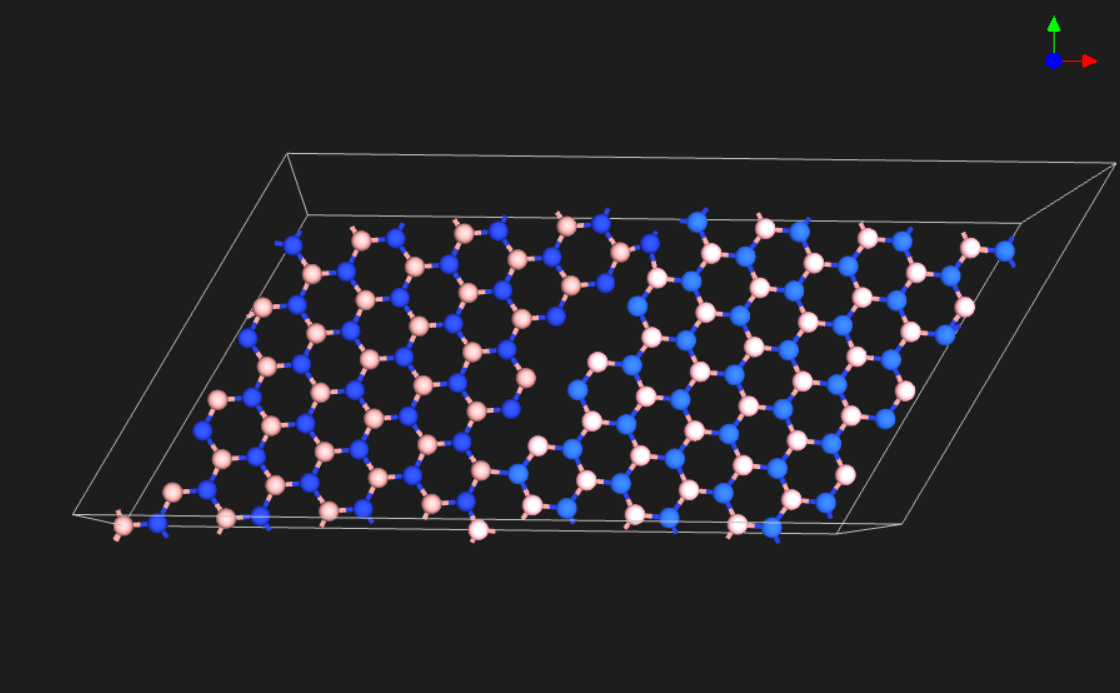} \\ \hline
            \texttt{D-0D-VAC} & Create point defect in a slab & Input: slab, target site, and removal operation.\newline Result: vacancy defect.\newline Example: vacancy in Graphene. & \includegraphics[width=\linewidth]{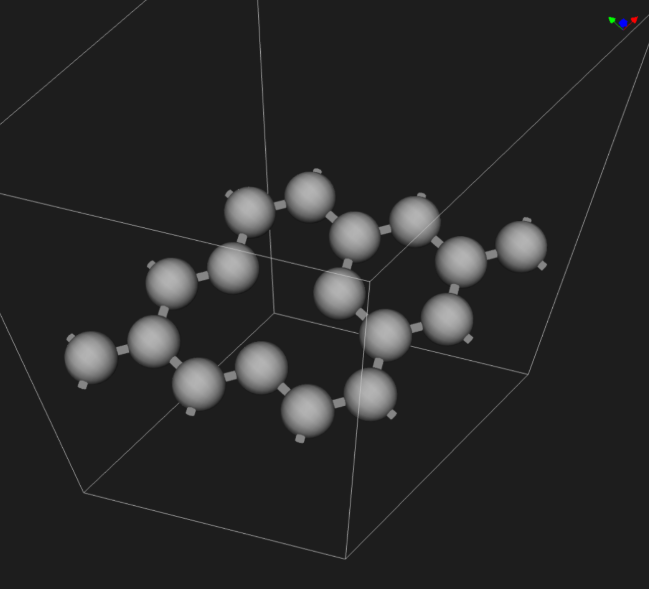} \\ \hline
            \texttt{D-0D-SUB} & Create point defect in a slab & Input: slab, target site, and replacement species.\newline Result: substitutional defect.\newline Example: Mg substitution in GaN. & \includegraphics[width=\linewidth]{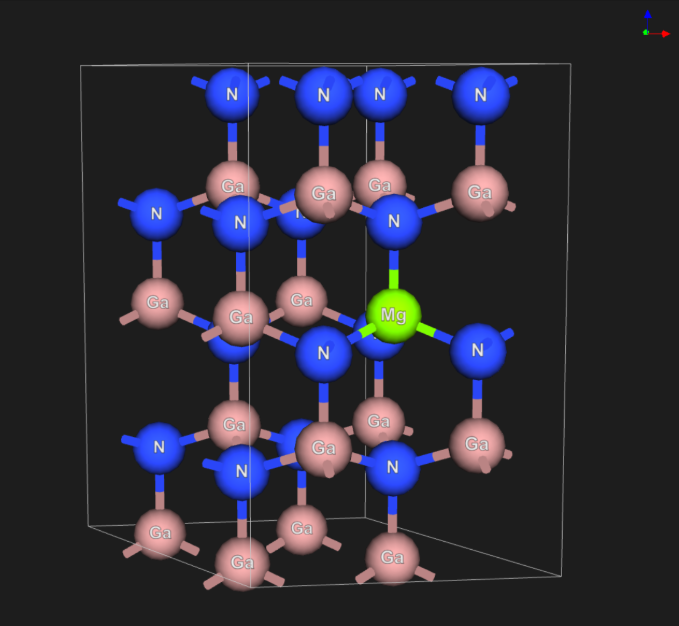} \\ \hline
            \texttt{D-0D-INT} & Create point defect in a slab & Input: slab, insertion site, and added species.\newline Result: interstitial defect.\newline Example: interstitial O in SnO. & \includegraphics[width=\linewidth]{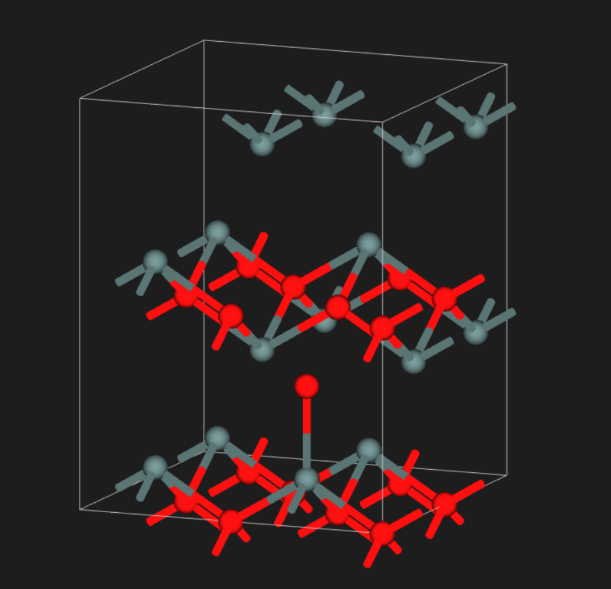} \\ \hline
            \texttt{X-2D-PAS} & Passivate slab & Input: slab, surface sites, and passivating species.\newline Result: passivated surface.\newline Example: H-passivated Cu(001) surface. & \includegraphics[width=\linewidth]{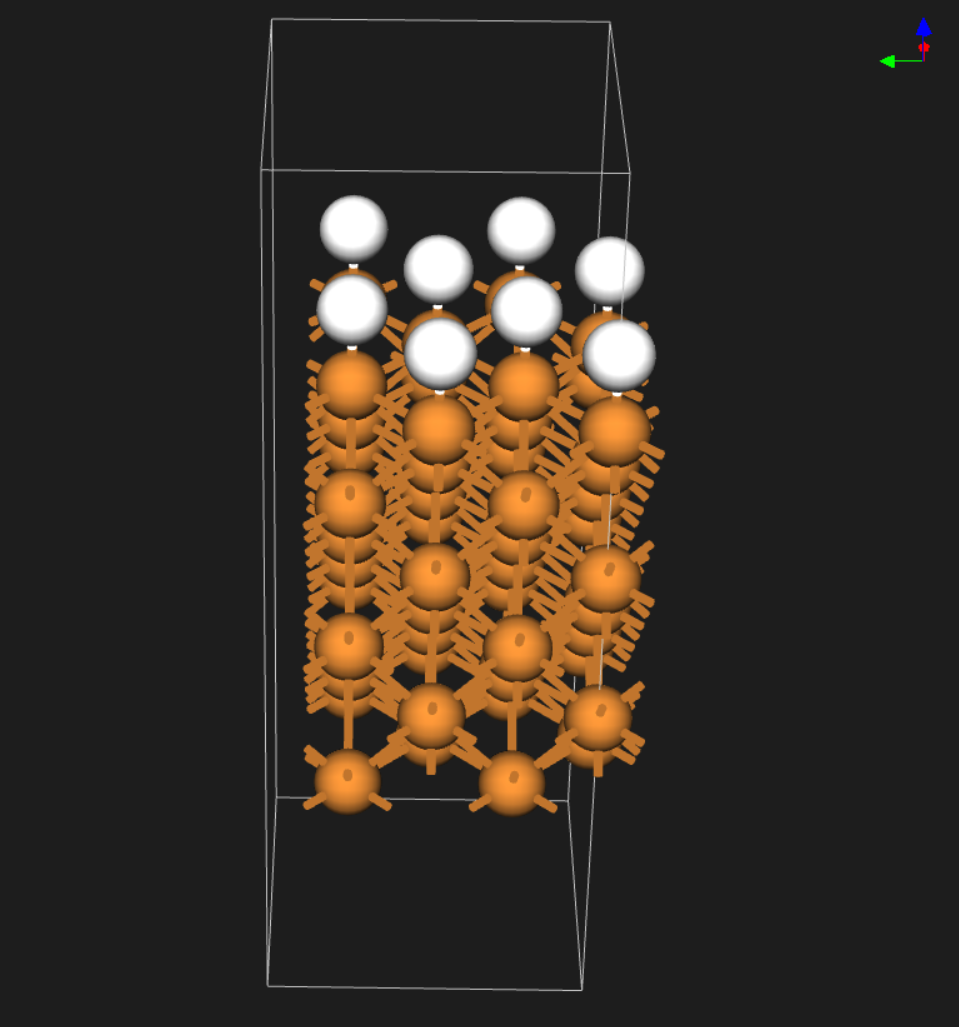} \\ \hline
            \texttt{X-2D-PER} & Add Sine Perturbation to a slab & Input: slab and perturbation amplitude or wavelength.\newline Result: distorted surface structure.\newline Example: Sine perturbation in Graphene sheet. & \includegraphics[width=\linewidth]{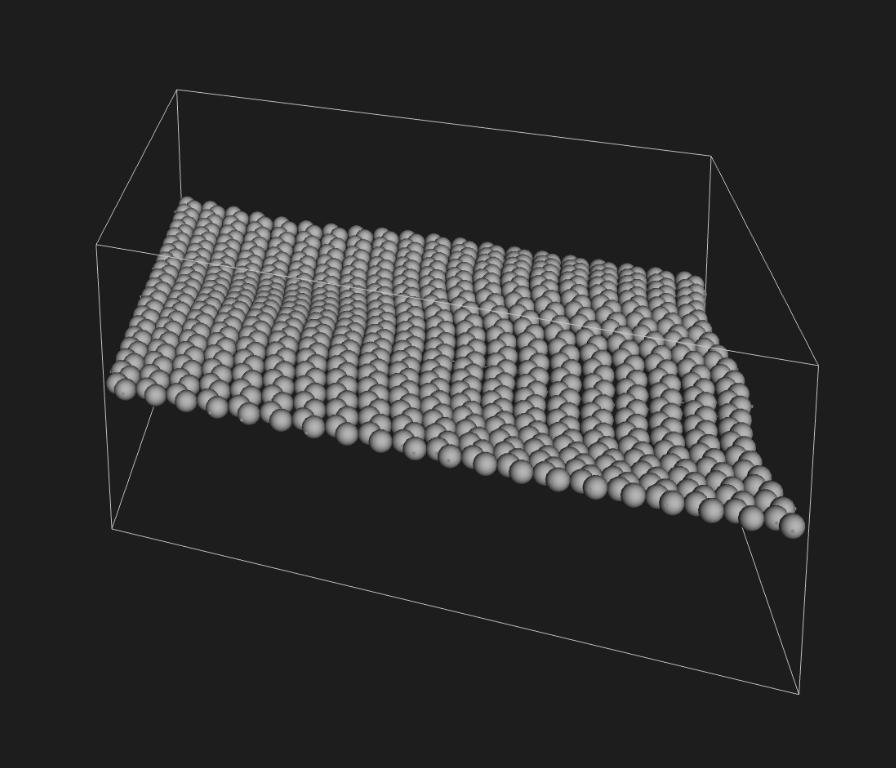} \\ \hline
            \texttt{X-1D-PAS} & Passivate nanoribbon & Input: nanoribbon edge sites and passivating species.\newline Result: edge-passivated ribbon.\newline Example: H-passivated graphene armchair nanoribbon. & \includegraphics[width=\linewidth]{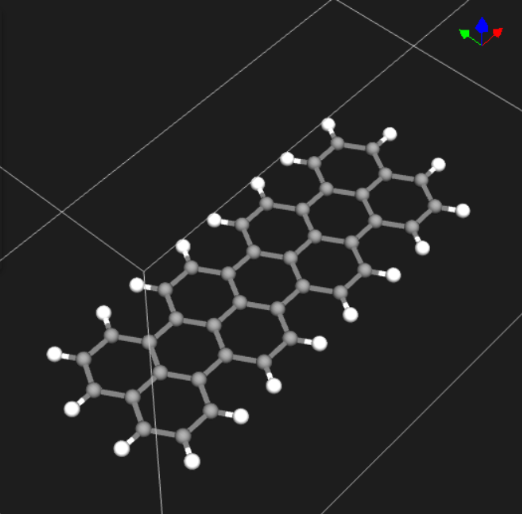} \\ \hline
            \texttt{X-0D-CUT} & Create box-cutout & Input: slab and box geometry for carving.\newline Result: finite cutout structure.\newline Example: box cutout in GaN(001) slab. & \includegraphics[width=\linewidth]{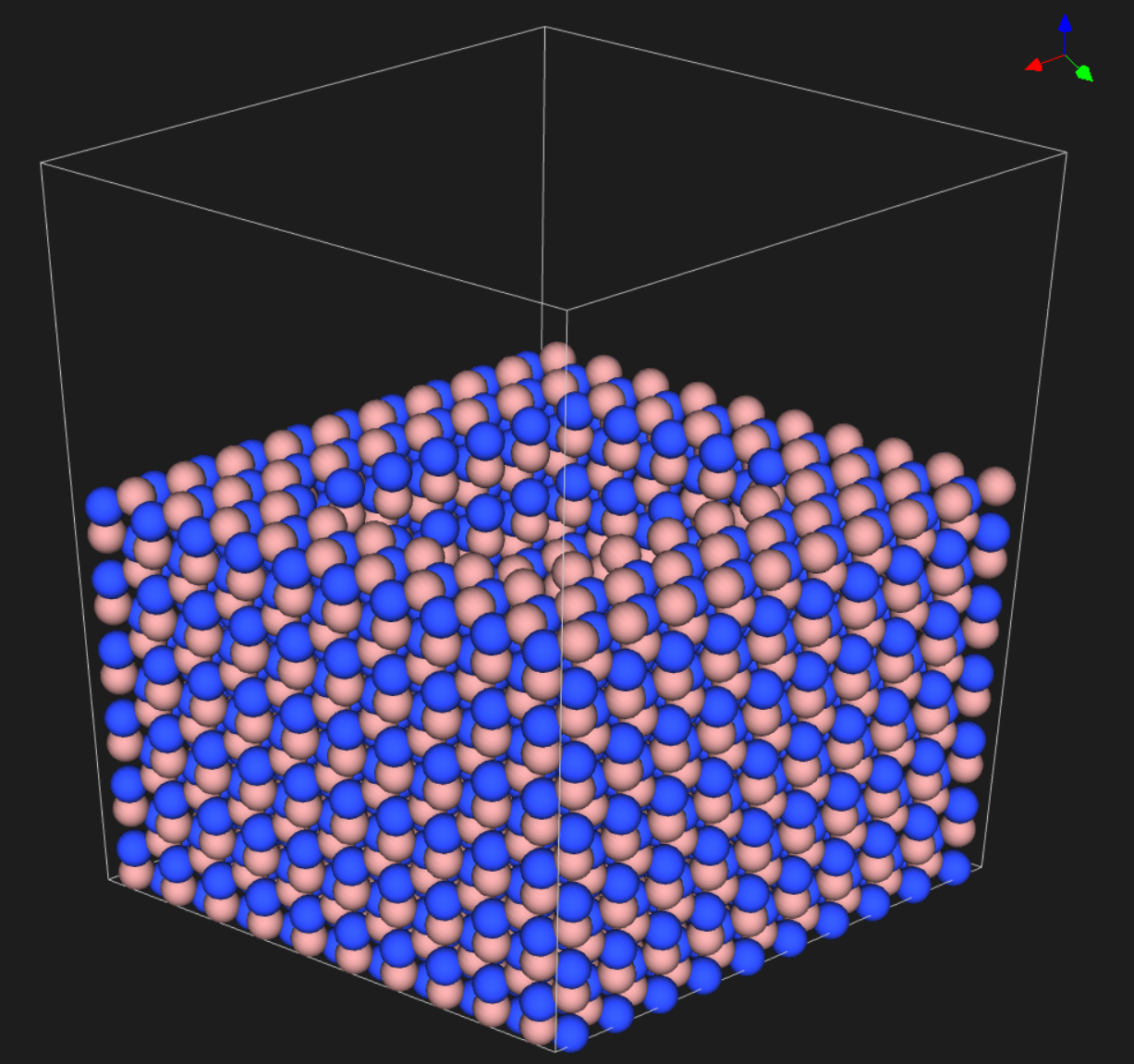} \\ \hline
        \end{longtable}
        \endgroup

    The same exact workflows can be executed in standard Python environments (JupyterLab, IPython) or directly in a web browser via Pyodide, lowering the setup and getting-started barriers while remaining interoperable with established toolkits such as pymatgen and ASE \cite{ong2013python, larsen2017atomic}. The web-based execution enables integration into web applications such as the Materials Designer platform, facilitating data exchange between Python and web/JavaScript environments.

    \section{Discussion}
\label{sec:discussion}

    Mat3ra-2D is motivated by a practical constraint in data-driven materials modeling: the structures most relevant to real devices and processes are rarely ideal bulk crystals.
    By making the generation of realistic slabs and interfaces explicit and reproducible, the framework supports the creation of training and benchmark datasets that better reflect the target deployment domain.

    \subsection{Procedure-oriented design and reuse}
        A central design choice in Mat3ra-2D is to treat realistic structure generation as a composition of reusable workflow steps rather than as a collection of isolated scripts. Slab generation, interface construction, defect insertion, passivation, and other operations are exposed through builders, analyzers, and reference data that can be combined in a controlled way. This has practical consequences for maintainability and extension: once a workflow for a representative case is defined, the same components can be reused across notebook examples, Python environments, and browser-based execution without redefining the concepts each time.

        This workflow-oriented structure also helps connect different layers of the ecosystem. The same transformation logic that appears in an educational notebook can be reused in a scripted dataset-generation workflow or embedded in a larger platform context. As the number of target structures grows, this reuse becomes increasingly important because it reduces duplication and keeps provenance handling, parameter conventions, and structure-building behavior consistent across examples.

    \subsection{Reproducibility and open access}
        A second contribution of Mat3ra-2D is the combination of provenance-aware generation with openly accessible execution environments. Provenance records capture the sequence of choices used to create a structure, including orientation, termination, thickness, strain, registry, and intermediate ranking criteria. This makes workflows inspectable and rerunnable rather than opaque. For realistic structures, where small modeling choices can produce meaningfully different systems, that transparency is essential.

        The notebook collections make this principle concrete. Generic notebooks expose individual transformations in a form that can be reused and adapted, while more advanced notebooks show how multiple transformations can be chained into longer structure-generation workflows. Because the examples are available through JupyterLite, they can be opened and executed in a web browser without local setup. This lowers the barrier to scrutiny and reuse and supports one of the central aims of the project: shared structure-generation workflows should be reproducible by others, not only by their original authors.

    \subsection{Implications for data and AI/ML}
        Dataset quality is shaped not only by the underlying electronic-structure method but also by how structures are defined, transformed, and curated. Many widely used datasets and benchmarks emphasize structures that are easy to enumerate and standardize, especially ideal bulk crystals, which can lead to models that perform well on in-distribution tests but fail to transfer to the heterogeneous and defect-rich structures that dominate electronic devices and processing conditions \cite{ward2016ml-framework, isayev2017ml-descriptors}. For 2D materials, relevant targets frequently involve slabs, reconstructions, point and extended defects, and film/substrate contacts, where behavior can be controlled by terminations, strain, stacking registry, and interface chemistry rather than by bulk stoichiometry alone \cite{haastrup2018computational, rasmussen2015computational}.

        Within this context, Mat3ra-2D helps make realistic structure generation part of the modeling problem rather than a hidden preprocessing step. When structures carry explicit provenance, datasets can be refined in a targeted way, category-specific failures can be diagnosed more systematically, and training or benchmark collections can be regenerated as assumptions evolve. In that sense, the framework complements broader efforts to assess first-principles reproducibility \cite{DFTReproducibility2016lejaeghere} and to connect ML behavior to interpretable materials representations \cite{EB-ARX-2023}. It also complements computational materials infrastructure that has accelerated high-throughput discovery through shared workflows and data platforms \cite{jain2013materialsproject, curtarolo2012aflowlib, saal2013openQMD, pizzi2016aiida, nomad}.

    \subsection{Practical considerations and limitations}
        The practical value of a framework such as Mat3ra-2D depends on how it is used to navigate an intrinsically large and uneven design space. Realistic 2D materials are combinatorial not only because there are many parent compounds, but also because each target structure introduces additional modeling choices. Interfaces require lattice matching, strain control, gap selection, and lateral registry decisions. Surfaces require choices of orientation, thickness, vacuum, and termination, and in some cases, reconstruction. Defect and disorder models introduce site selection, concentration, chemical identity, and symmetry considerations. Even when every transformation is individually well defined, the number of plausible combinations can grow rapidly.

        This clarifies the scope of the present work. Mat3ra-2D provides machinery to define, build, annotate, and share realistic structures, but it does not remove the scientific judgment required to decide which structures are representative, which intermediate candidates should be selected, or which fidelity level is appropriate for downstream simulation. Reproducible generation is not the same as validated generation: a structure can be regenerated exactly and still be a poor approximation to the experimental system if the wrong termination, reconstruction, or defect model was chosen. The framework, therefore, improves inspectability and reuse, but it does not replace benchmark design, convergence studies, or comparison with experiment.

    \subsection{Open-source development and interoperability}
        Mat3ra-2D is intended to complement, rather than replace, existing computational materials infrastructure. Data platforms and workflow systems have already enabled large-scale computation and data exchange, and the present framework focuses on the realistic structure-generation layer that connects input concepts, workflow steps, and reusable examples. This positioning is important for adoption: practical use will depend on robust converters, validators, database connectors, and reference-data pipelines that streamline the path from structure generation to calculation, storage, and reuse.

        The open-source ecosystem is equally important. Schemas, reference data, code, and notebooks are most useful when they can be inspected, adapted, and extended by others. Open access to the notebooks is particularly valuable because it allows users to move from reading about a workflow to running it directly in the browser or reusing it in a conventional Python environment. In this sense, openness is not only a distribution choice; it is part of the method by which workflows become auditable, reusable, and easier to integrate into other toolchains.

    \subsection{Future directions}
        Several directions are important for realizing the full potential of realistic structure generation with provenance. One is validation: systematically generated interfaces, surfaces, and defect models should be compared against experimental observations and trusted computational benchmarks to identify which classes of approximations are reliable and which require refinement. Another is scalable dataset design: because realistic structure spaces are too large to enumerate exhaustively, useful progress will likely come from benchmark suites, agreed reporting conventions, and selective sampling strategies rather than brute-force coverage.

        Broader access through open notebooks and browser execution also creates opportunities beyond benchmarking. Educational examples, surface-specific property datasets, transfer-learning studies from bulk to low-dimensional systems, and inverse-design loops in which target properties guide structure generation all become easier to organize when realistic structures can be generated, inspected, and regenerated through shared workflows. The long-term value of Mat3ra-2D will depend on how well these future directions combine accessibility with scientific rigor, allowing realistic structure generation to become both easier to use and easier to trust.
    \section{Conclusion}
\label{sec:conclusion}

    This work presented Mat3ra-2D as a framework for constructing realistic 2D materials and related structures with explicit, machine-readable provenance.
    Building on the M-CODE ontology and categorization framework \cite{biryukov2026mcode}, we illustrated representative outcomes and workflows for interfaces and other low-dimensional structures, and introduced notebook-based examples that support reproducible structure generation.
    A central idea of the project is that published structures should be reproducible by anyone from openly available online notebooks.
    By prioritizing realism, provenance, accessibility, and reuse, Mat3ra-2D aims to enable dataset design and modeling workflows that better reflect device-relevant materials complexity and improve the practical utility of AI/ML models.

    \section*{Acknowledgements}
    This work was supported in part by NIST 70NANB24H205. We used large language models (LLMs) to assist with drafting and editing portions of this manuscript. All content was reviewed and edited by the authors, who take full responsibility for the final text.
    
    \section*{Data and Software availability}
    M-CODE and the associated schemas are maintained as open-source data standards. The reference implementation is accessible as a package on GitHub and distributed via PyPI as \texttt{mat3ra-esse}. Related ecosystem packages and notebooks are available on GitHub and PyPI, including \texttt{mat3ra-standata}, \texttt{mat3ra-made}, and \texttt{mat3ra-api-examples} \cite{mat3ra-esse-pypi-package, mat3ra-standata-pypi-package, mat3ra-made-pypi-package, mat3ra-api-examples-pypi-package}.

    \bibliographystyle{unsrtnat}
    \bibliography{references}

\end{document}